\documentclass[a4paper,superscriptaddress,onecolumn,preprint,pre,showpacs,amsmath,amssymb,floatfix]{revtex4-1}
\usepackage{amsfonts}
\usepackage{amssymb}
\usepackage{amsmath}
\usepackage{graphicx}
\usepackage{pgfplots}
\usepackage{etex}
 \usepackage{morefloats}
\usepackage{mathbbol} 

\def\kt{\mathfrak{t}}
\def\kh{\mathfrak{h}}

\def\km{\mathfrak{m}}
\def\kT{\mathfrak{T}}
\def\kH{\mathfrak{H}}

\begin{document}
\title{Statistical properties of structured random matrices}
\author{Eugene Bogomolny}
\affiliation{Universit\'e  Paris-Saclay, CNRS, LPTMS, 91405 Orsay, France}   
\author{Olivier Giraud}
\affiliation{Universit\'e  Paris-Saclay, CNRS, LPTMS, 91405 Orsay, France}  
\date{December 21, 2020}
\begin{abstract}
Spectral properties of Hermitian Toeplitz, Hankel, and Toeplitz-plus-Hankel random matrices with independent identically distributed entries are investigated. Combining numerical and analytic arguments it is demonstrated that spectral statistics of all these random matrices is of intermediate type, characterized by (i) level repulsion at small distances, (ii) an exponential decrease of the nearest-neighbor distributions at large distances, (iii) a non-trivial value of the spectral compressibility, and (iv) the existence of non-trivial fractal dimensions of eigenvectors in Fourier space.  Our findings show that intermediate-type statistics is more ubiquitous and universal than was considered so far and open a new direction in random matrix theory.   
\end{abstract}

\maketitle

\section{Introduction}

Matrices are omnipresent structures in extremely varied branches of physics and mathematics and there exist many different types of matrices tailored for specific problems. To get a certain overview of  a general matrix classification one could order matrices  according to  their complexity. The well-known Kolmogorov complexity of a string is the minimal length of a program which can calculate the string.  A possible way  to  measure the (arithmetic) complexity of a matrix is by counting the minimal number of operations needed to perform certain non-trivial operations, e.g., to find the inverse matrix, or matrix eigenvalues, with a given precision (see e.g.~\cite{stoc, coppersmith, legall} and references therein).  

For generic  $N\times N$ matrices, standard algorithms  (like the Gauss-Jordan  elimination) require
 $\mathcal{O}(N^3)$ operations, though  more refined algorithms  reduce it (up to logarithmic corrections)  to $\mathcal{O}(N^{\omega})$ with $\omega\approx 2.3728639$ \cite{legall}.  Nevertheless there exist special types of matrices of  lower complexity which generically  necessitate smaller number of operations. 
 
The most investigated  class of such low-complexity matrices consists of matrices with small-rank displacement structure \cite{kailath_1, kailath_2}. They are characterized by the existence of a linear operator  that transforms all matrices from the class into matrices of small rank. Two main types of such displacement operators were used, the Toeplitz-like displacement operator $\nabla_{A,B}(M)=M-A\,M\,B$ and the Hankel-like displacement operator $\Delta_{A,B}(M)=A\,M-M\,B$, where $A$ and $B$ are arbitrary matrices. The rank of matrices $\nabla_{A,B}(M)$ and $\Delta_{A,B}(M)$ depends on the choice of $A$ and $B$; the minimal rank, $r$, is called the displacement rank of $M$.  A matrix is referred to as a structured matrix if its displacement rank is much smaller than its dimension.  
The importance of this notion comes from the theorem proved in \cite{kailath_1} 
that the standard $\mathcal{O}(N^3)$ number of operations  needed, e.g.~,  
to inverse a matrix can be replaced by 
$\mathcal{O}(r\,N^2)$. By using more sophisticated algorithms, this number can even be reduced to  $\mathcal{O}(r\, N\, \ln N)$  \cite{morf, heinig, pan}.     

The best-known examples of structured matrices are Toeplitz ($T_{mn}$), Hankel ($H_{mn}$), and Toeplitz-plus-Hankel ($(T+H)_{mn}$) matrices, whose matrix elements have  the  following form
\begin{equation}
T_{mn}=t_{m-n} ,\quad H_{mn}=h_{m+n},\quad (T+H)_{mn}=t_{m-n}+h_{m+n}
\label{main_matrices}
\end{equation}
with $m,n=1,\ldots,N$ and $t_i$, $h_j$ arbitrary real or complex numbers.
The matrices considered in \eqref{main_matrices} have a long history : Hankel matrices were introduced in 1861 \cite{hankel1861} and Toeplitz matrices in 1911 \cite{toeplitz1911}. They appear naturally in various fields of mathematics and physics such as differential and integral equations, functional analysis, probability theory, statistics, numerical analysis, theory of stationary processes, signal and image processing, control theory, integrable models,  among many others (see, e.g.,  \cite{grenander, iovidov, bottcher, peller, deift1, deift2, krasovsky} and references therein). The existence of  algorithms inverting these matrices in $\mathcal{O}(N^2)$ operations were known for a long time \cite{levinson, trench1, trench2, trenchhankel, heinig}.  Examples of displacement structures for these matrices is briefly discussed in Appendix~\ref{displacement}. 
Considerable efforts were done to find the asymptotic behavior of the determinants and eigen-problems for  matrices \eqref{main_matrices} in the limit of  large matrix dimensions (see, e.g., \cite{krasovsky, deift1, deift2} and references therein).  It appeared that all these calculations require additional regularity conditions of matrix elements (e.g., a finite number of Fisher-Hartwig singularities). Very irregular matrices, that is, without any particular structure other than \eqref{main_matrices}, seem to be inaccessible to known analytic methods. 

The investigation of irregular Hermitian Toeplitz matrices was initiated in \cite{bogomolny}, where elements $t_k$ were taken as independent and identically distributed  (i.i.d.) random variables (with $t_{-k}=t_k^{*}$). A central aspect of the study of random matrices is the investigation of their statistical spectral properties. From the above-mentioned fact that Toeplitz matrices are low-complexity matrices it seems natural that their spectral statistics differ from the Wigner-Dyson statistics of usual random matrix ensembles used to describe chaotic systems \cite{bohigas}. It was shown in \cite{bogomolny} that spectral statistics of random Toeplitz matrices is of intermediate type, which is characterized by level repulsion, as for usual random matrix ensembles \cite{mehta}, but with exponential decrease of nearest-neighbor spacing distributions, as for the Poisson distribution typical for integrable models \cite{berry}. Such a type of intermediate spectral statistics was first observed in the Anderson model at the point of metal-insulator transition \cite{shklovskii, altshuler}, and later in certain pseudo-integrable billiards \cite{schmit, wiersig} and quantum maps \cite{GirMar04}.  More precisely, \cite{bogomolny} showed that spectral statistics of random Toeplitz matrices are  well described by the semi-Poisson distribution, which is the simplest model where only the nearest-neighbor levels interact (an approach described in \cite{gerland}). 
 
The main purpose of this paper is to investigate  statistical properties  of other structured matrices beyond the Toeplitz class, namely random  Hermitian Hankel and Toeplitz-plus-Hankel matrices where elements $t_k$ and $h_k$ in \eqref{main_matrices}  are i.i.d.~Gaussian random variables with zero mean and unit variance (for complex elements real and imaginary parts are i.i.d.~standard Gaussian random variables).  
The main conclusion of the paper is that spectral statistics of all these low-complexity matrices is of intermediate type and well described by a gamma distribution. Moreover, eigenvectors of these matrices are multifractal in Fourier space, which is typical for models with intermediate statistics. Such a multifractal behavior of eigenstates was identified in wavefunctions of the Anderson model at metal-insulator transition \cite{Weg80, Aok83, CasPel86} and in certain random matrix ensembles \cite{BogGir11}. Quite remarkably, as we show here, such features are also present in models as simple as random Toeplitz or Hankel matrix ensembles.

The plan of the paper is the following. Section~\ref{new_method} is devoted to the investigation of a simple heuristic method which permits to obtain  explicit approximate formulas for spectral statistics of structured matrices. These results are then applied  to random Hermitian  Toeplitz, Hankel, and Toeplitz-plus-Hankel matrices  with independent matrix elements. The results demonstrate the intermediate character of spectral statistics for these matrices and confirm the fact, observed in \cite{bogomolny}, that random Toeplitz matrices are well approximated by the semi-Poisson distribution. Other functions characterizing the spectrum are discussed in Section~\ref{other_spectral_properties}.
In Section~\ref{numerical_correlation_functions} it is  demonstrated that  the results of direct large-scale numerical calculations for different correlation functions for the above matrices agree well with the obtained approximate formulas. The summary of the obtained results is done in Section~\ref{conclusion}.  In Appendix~\ref{displacement} the simplest displacement structures for the considered matrices are briefly discussed. Appendix~\ref{plasma} is devoted to the construction of short-range plasma models which have the same power-low behavior as Hankel and Toeplitz-plus-Hankel matrices discussed in the main text. 

\section{Wigner-type approximate formulas}\label{new_method}

The purpose of the present Section is to  obtain heuristically  simple approximate formulas for structured random matrices. Our guiding principle is the construction of the Wigner-type surmises for nearest-neighbor distributions in standard Wigner-Dyson ensembles of random matrices.

\subsection{Wigner-Dyson ensembles}\label{wdens}

The usual Wigner-Dyson ensembles of random matrices are the Gaussian orthogonal ensemble (GOE), the Gaussian unitary ensemble (GUE), and the Gaussian symplectic ensemble (GSE) characterized, respectively, by the Dyson index 
$\beta=1,2,4$.
For these ensembles, it is well-known (see, e.g., \cite{mehta, porter}) that the nearest-neighbor spacing distribution $P_0(s)$ (i.e., the probability that two levels are separated by a distance $s$ with no level inbetween) is well approximated  by the Wigner surmise 
\begin{equation}
\label{p0sRMT}
P_0(s)=a(\beta) s^{\beta} \mathrm{e}^{-b(\beta) s^2}\,,
\end{equation}
with constants $a(\beta)$ and $b(\beta)$ determined from the normalization conditions
\begin{equation}
\int_0^{\infty}P_0(s)\mathrm{d}s=1,\qquad \int_0^{\infty}sP_0(s)\mathrm{d}s=1\ .
\end{equation} 
The success of such a surmise is based on the simple fact that any function which has the correct behavior $\sim s^\beta$ at small values of the argument and is quickly decreasing  at large values of the argument should be a reasonably good approximation for the true function, provided normalization fixes the otherwise arbitrary parameters $a(\beta)$ and $b(\beta)$. Of course, deviations between the exact result and the simple expression \eqref{p0sRMT} do exist, but as  the function at  large argument is small, they are practically unobserved. The accuracy of the Wigner surmise is so high that the exact function, given by a solution of a certain Painlev\'e equation \cite{jimbo, mehta}, is very rarely used, mainly   in  monumental calculations of the Riemann zeta function \cite{odlysko}. In most other cases the Wigner surmise is sufficient. 

\begin{figure}
\begin{center}
\includegraphics[width=.4\linewidth]{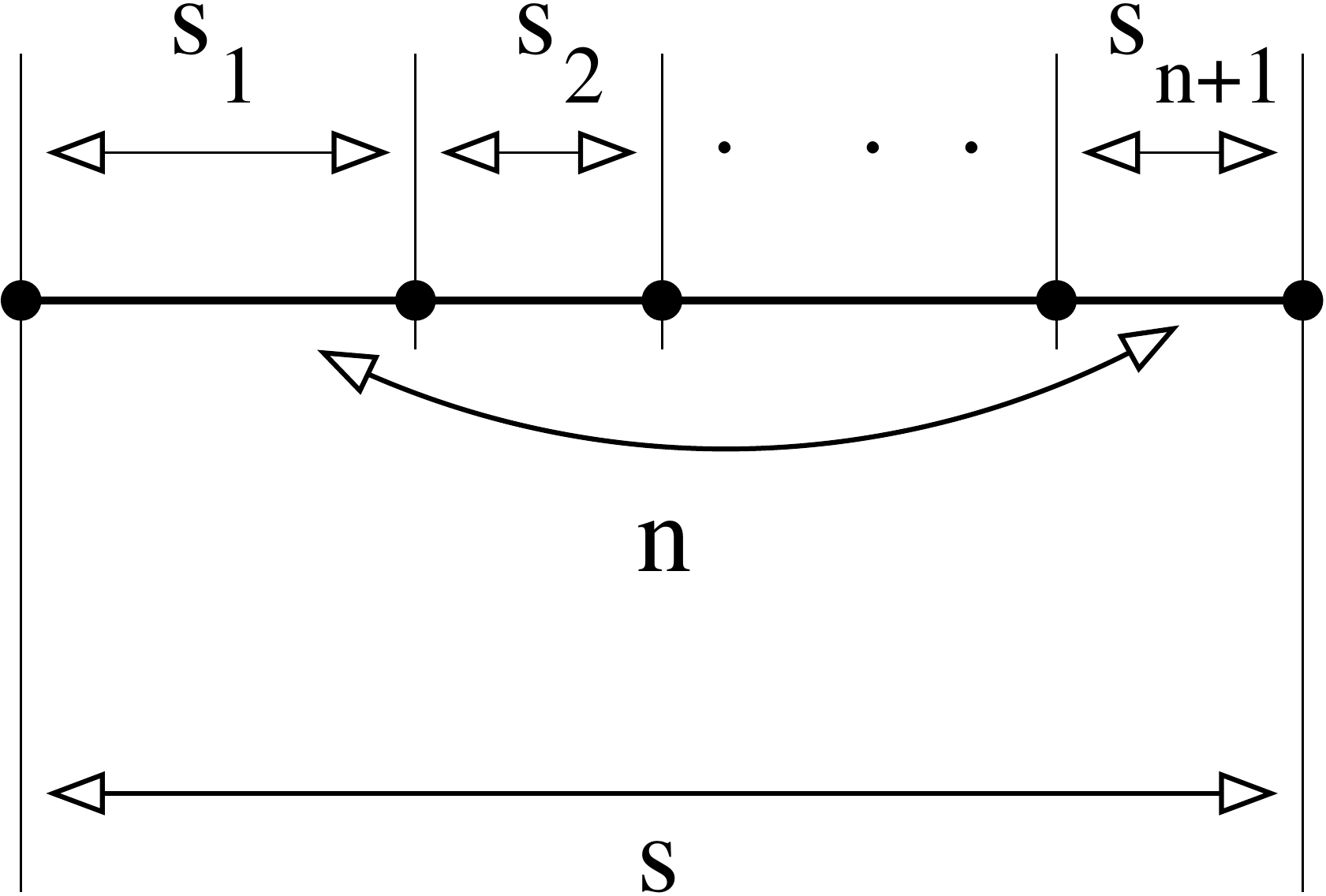}
\end{center}
\caption{Configuration of eigenvalues for $P_n(s)$ correlation function.}
\label{eigenvalues}
\end{figure}

Much less used are the analogous Wigner-type surmises for the higher-order nearest-neighbor spacing distributions $P_n(s)$, which are the probabilities that two eigenvalues are separated by a distance $s$ with exactly $n$ eigenvalues between them. Again, the accuracy of such an approximate formula for $P_n(s)$ mainly depends on the correctness of the small-$s$ behavior and of the quick asymptotic decrease of the function at large argument. The main ingredient to obtain these surmises is therefore to determine the small-argument behavior of $P_n(s)$.  This can  be readily obtained from the exact joint eigenvalue probability density, which for the Wigner-Dyson random matrix ensembles is known to be \cite{mehta} 
\begin{equation}
P(e_1,\ldots, e_N)\sim \prod_{1\leq i<j\leq N} |e_i-e_j|^{\beta} \prod_{k=1}^N \mathrm{e}^{-V(e_k)}
\label{exact_proba}
\end{equation} 
where $V(e)$ is a confining potential and $\beta=1,2,4$. 
The matrices that contribute most to  $P_n(s)$ for $s\sim 0$ are those for which $n+2$ eigenvalues are close to a certain value $\lambda$ and thus almost degenerate, with a distance $s\ll 1$ between the largest and the smallest of these eigenvalues (see Fig.~\ref{eigenvalues}). If we assume that all other eigenvalues are separated from $\lambda$ by a gap $\gg s$, then the product $\prod_{i<j}$ over all pairs of eigenvalues in \eqref{exact_proba} splits into two parts: a first product $\Pi_1$ involving only those eigenvalues that are close to $\lambda$, and a second product $\Pi_2$ including all the other terms. In the product $\Pi_2$, eigenvalues close to $\lambda$ are paired with eigenvalues far from $\lambda$, and thus can be replaced by $\lambda$. Under this approximation, only the product $\Pi_1$ will contribute to the small-$s$ behavior of $P_n(s)$. This product $\Pi_1$ can be expressed solely in terms of spacings $s_1,...,s_{n+1}$ between consecutive eigenvalues close to $\lambda$ (see  Fig.~\ref{eigenvalues}). At small $s$ the confining potential can be discarded, and we thus have
\begin{equation}
\label{pnsrmt}
P_n(s)\underset{s\to 0}{\sim}\int_0^{\infty} \mathrm{d} s_1 \mathrm{d} s_2 \ldots  \mathrm{d} s_{n+1}\prod_{k=0}^n\prod_{i=1}^{n-k+1}\left(s_i+\cdots+s_{i+k}\right)^{\beta} \delta\left(s-\sum_{k=1}^{n+1} s_k\right)\,.
\end{equation}
Substituting $s_k=s y_k$ in \eqref{pnsrmt} one can extract the leading power of $s$ by merely counting the different contributions (the delta function accounting for a -1 contribution); the remaining integral just gives an overall multiplicative constant. One directly gets
\begin{equation}
P_n(s)\underset{s\to 0}{\sim} s^{\gamma_n},\qquad  \gamma_n=\beta\frac{(n+2)(n+1)}{2}+n\ .
\label{gamma_rm}
\end{equation}
Assuming, as in the  usual Wigner surmise, that all correlation functions have Gaussian decay at large argument  one gets an approximate formula for $P_n(s)$ of the form
\begin{equation}
P_n(s)=a_n s^{\gamma_n}\exp (-b_n s^2)\,,
\label{W_D}
\end{equation}
where constants $a_n$ and $b_n$ are calculated from the standard normalization conditions 
\begin{equation}
\int_0^{\infty}P_n(s)\mathrm{d}s=1,\qquad \int_0^{\infty}sP_n(s)\mathrm{d}s=n+1\ .
\label{normalization_ps}
\end{equation}
These formulas are not new and have been derived  in, e.g., \cite{rao} and (from  different considerations) in \cite{abul_magd}.

The main drawback of the above approach is that it requires the knowledge of the exact eigenvalue distribution. Now we shall obtain the same result from simple  arguments, without using the specific form \eqref{exact_proba}, and extend it to more general situations.

\subsection{Wigner-Dyson revisited and intermediate-type statistics}\label{wdensrev}
Suppose, as above, that the small-$s$ behavior of $P_n(s)$ comes from matrices with $n+2$ eigenvalues close to a certain fixed value $\lambda$ and all other eigenvalues far away from $\lambda$. This means that only eigenvalues close to $\lambda$ contribute to $P_n(s)$ as $s\to 0$, and the other eigenvalues can be ignored. One can then restrict oneself to a $(n+2)\times (n+2)$ Hermitian matrix $M_{ij}$ whose eigenvalues are almost degenerate, that is
\begin{equation}
M_{ij}=\lambda\delta_{ij}+\epsilon \km_{ij} ,\qquad |\epsilon|\ll 1 .
\label{close_to_identity}
\end{equation}
We are looking for the number of independent variables in $\km$ whose non-zero values lift the degeneracy. Since that matrix is Hermitian, the total number of independent matrix elements  is $N_t=n+2+\beta \binom{n+2}{2}$, with $\beta=1$ for real and $\beta=2$ for complex matrices. But adding a constant value to all $\km_{ii}$ does not change the relative position of eigenvalues, so that this number has to be diminished by $1$ (one could, e.g., impose that $\km_{11}=0$). Therefore we get 
\begin{equation}
q_n=n+1+\beta \binom{n+2}{2}
\label{q_n_WD}
\end{equation}
 variables whose non-zero values lift the degeneracy. 

Let  $e_j$ be eigenvalues of $M_{ij}$. Setting $e_j=\lambda+\epsilon v_j$ for $j=1,\ldots,n+2$, the $v_j$ are  solutions of $\det(v\,\delta_{ij}-\km_{ij})=0$. When $\km_{ij}=\mathcal{O}(1)$  all $e_j$ are, in general,  different and are of the order of $\epsilon$ around $\lambda$.  
For the considered ensembles all $q_n$ matrix elements are independent random variables with a certain non-zero probability density. The spacing distribution is then obtained by a $q_n$-fold integral over these variables. The probability that all eigenvalues of matrix $M_{ij}$  are within a small distance $s$ from $\lambda$ is the probability of the event that all the $q_n$ variables are of the order of $s$. One can therefore rescale each variable in the $q_n$-fold integral by $s$, which gives a probability proportional to $s^{q_n}$, and thus
\begin{equation}
P(|e_j-\lambda|<s) \underset{s\to 0}{\sim} s^{q_n} .
\end{equation}
This quantity corresponds to the cumulative spacing distribution $\int_0^s P_n(y)dy$. Therefore the spacing distribution $P_n(s)$ has the following  limiting value  
\begin{equation}
\label{pnsRMT}
P_n(s)\underset{s\to 0}{\sim} s^{\gamma_n}, \qquad \gamma_n=q_n-1.
\end{equation}
For the Wigner-Dyson ensembles $q_n$ is given by \eqref{q_n_WD}, thus this expression  agrees with the above result \eqref{gamma_rm} obtained from the exact joint distribution. The behavior of $P_n(s)$ at large values of the argument is then determined by the fact that each eigenvalue interacts with all other eigenvalues, which suggests  the quadratic exponent in \eqref{W_D}

Mathematically, the number $q_n$, known as the codimension, determines the minimal number of independent parameters in a Hermitian matrix that has $n+2$ degenerate eigenvalues. The well-known theorem of von Neumann and Wigner \cite{Neumann, keller} states that the codimension of Hermitian matrices with $k$ degenerate eigenvalues is $k^2-1$ for complex matrices and $\frac12(k+2)(k+1)$ for real ones. When $k=n+2$ these expressions agree with $q_n$ in Eq.~\eqref{q_n_WD} obtained from matrices of size $n+2$. 

We are unaware of exact results about codimensions for matrix families considered in the paper. 
Nevertheless, the codimensions of any sub-class of Hermitian matrices are independent of matrix dimensions, as they are upper-bounded by the above size-independent values. It is then natural to conjecture that they can be determined by considering the smallest possible matrix with $n+2$ degenerate eigenvalues. The above discussion shows that such an approach works well for the Wigner-Dyson ensembles. Below it will be applied to Hermitian  Toeplitz, Hankel, and Toeplitz-plus-Hankel  ensembles of random matrices.

In order to get a complete Wigner-like surmise we must additionally fix the behavior at large argument. In the case of the intermediate-type ensembles considered in the present paper one can argue \cite{shklovskii, altshuler, schmit} that the interaction between eigenvalues has to be of short range and asymptotically the nearest-neighbor distributions decrease only as an exponential of the distance between eigenvalues (which is typical  in short-range interaction thermodynamics). Combining both asymptotic behaviors at large and small $s$, one gets that the Wigner-like surmise for intermediate-type matrices should be of the form
\begin{equation}
P_n(s)=a_n s^{\gamma_n}\exp(-b_n s),
\label{gamma_pdf}
\end{equation}
where $a_n$ and $b_n$ are fixed by the normalization conditions \eqref{normalization_ps}, that is,
\begin{equation}
a_n=\frac{1}{\Gamma(\gamma_n+1)}\Big (\frac{\gamma_n+1}{n+1}\Big )^{\gamma_n+1} ,\qquad b_n=\frac{\gamma_n+1}{n+1}.
\label{correct_norm}
\end{equation}
The distribution \eqref{gamma_pdf} belongs to the family of gamma distributions. It is determined by the quantity $\gamma_n=q_n-1$ only, with $q_n$ being the minimal  number of independent matrix elements in a small vicinity of a degenerate matrix such that any variations of them lift the degeneracies of matrix eigenvalues.  Another definition of $q_n$ is that it is equal to the total number of parameters minus the number of 'zero modes', that is, the number of parameters whose variation does not remove the eigenvalue degeneracy.

\subsection{Toeplitz matrices}\label{wdensrevT}

Let us now consider a random $(n+2)\times (n+2)$ Hermitian Toeplitz matrix $T_{jk}=t_{j-k}$, $1\leq j,k\leq n+2$, where $t_i$ are real or complex i.i.d.~Gaussian random variables.  A Hermitian Toeplitz matrix of size $n+2$ has $n+1$ distinct off-diagonal elements, and a single real diagonal entry. In total this gives $N_t=1+\beta(n+1)$ independent variables, where $\beta=1$ for real and $\beta=2$ for complex matrices.  

If a degenerate matrix is perturbed by a Toeplitz matrix $\kT$ in such a way that the perturbation does not lift the degeneracy, then $\kT$ has to be proportional to the identity matrix, $\kT=\lambda \mathbb{1}$. There is only one such matrix, with $\kt_0=\lambda$ and all other elements zero.  It implies that there exists  only one 'zero mode'. Therefore $q_n=N_t-1=\beta(n+1)$ and, consequently  
\begin{equation}
\gamma_n=\beta (n+1)-1\ . 
\label{gamma_Toeplitz}
\end{equation}
The gamma distribution \eqref{gamma_pdf} with such $\gamma_n$ and  $\beta=1$ corresponds to  the Poisson distribution, while for $\beta=2$ it  coincides with the semi-Poisson distribution, which agrees with the results of \cite{bogomolny} for random Toeplitz matrices.  

\subsection{Special Toeplitz-plus-Hankel matrices}\label{wdensrevTH}
We now apply the same method to Hermitian Toeplitz-plus-Hankel matrices having  the form $t_{i-j}+h_{i+j}$. Elements $h_i$ may either be independent from the elements $t_j$ or depend on them. Let us start with the second situation, where entries of the Hankel matrix are given in terms of the $t_j$. This situation arises when considering the spectrum of a real symmetric Toeplitz matrix. Indeed, it is well-known that the spectrum of such a matrix can be split into two sets of eigenvalues, the so-called reciprocal and anti-reciprocal sets, associated respectively with symmetric and skew-symmetric eigenvectors \cite{andrew, cantori}. These sets (for real Toeplitz matrices of even dimension) are given by eigenvalues of matrices of the form
\begin{equation}
(T +\eta  H)_{ij}=t_{|i-j|}+\eta\, t_{i+j-1}\,,
\label{even}
\end{equation}
where $\eta=\pm 1$ and $t_i$ are real i.i.d.~Gaussian random variables. 

Any $(n+2)\times(n+2)$ matrix of the form \eqref{even} is determined by the $N_t=2n+4$ elements $t_0,\ldots, t_{2n+3}$. Once again, the identity matrix belongs to this ensemble, thus the number of 'zero modes' corresponds to the number of parameters for which a matrix \eqref{even} is  proportional to the identity,
\begin{equation}
\kT+\eta\,\kH=\lambda\mathbb{1}. 
\end{equation}
 Such a matrix has only two free parameters. Indeed, it is subjected to the restrictions $\kt_{ij}+\eta\,\kh_{ij}=\lambda \delta_{ij}$, yielding for $\kt_{ij}=\kt_{|i-j|}$ and $\kh_{ij}=\kt_{i+j-1}$,
\begin{align}
\label{off_diag0}
\kt_0+\eta\, \kt_{2i-1}&=\lambda,\qquad i=1,\ldots, n+2,\\
\kt_{i-j}+\eta\, \kt_{i+j-1}&=0,\qquad  i=j+1,\ldots,n+2. 
\label{off_diag}
\end{align}
Condition \eqref{off_diag0} is equivalent to 
\begin{equation}
\eta \kt_{2i-1}=\lambda-\kt_0,\qquad i=1,\ldots, n+2,
 \label{t_odd}
\end{equation}
which fixes all odd coefficients in terms of $\lambda$ and $\kt_0$, while condition \eqref{off_diag} taken at $i=j+1$ gives 
\begin{equation}
\kt_{2j}=-\eta\, \kt_1 ,\qquad j=1,\ldots, n+1,
\label{t_even}
\end{equation}
which fixes all even coefficients apart from $\kt_0$. The two remaining free parameters are thus $\kt_0$ and $\lambda$, yielding two 'zero modes'. The number of independent parameters minus the number of zero modes is then $q_n=N_t-2=2n+2$, and one gets for these matrices 
\begin{equation}
\gamma_n=2n+1,
\label{2n1}
\end{equation}
which corresponds to the semi-Poisson distribution. Again, this is in agreement with the findings in \cite{bogomolny}.

\subsection{Independent Toeplitz-plus-Hankel matrices}\label{wdensrevTHind}
Suppose now that coefficients of the Toeplitz and Hankel matrices are fully independent. Namely, we consider Toeplitz-plus-Hankel matrices of the form
\begin{equation}
(T+H)_{ij}= t_{i-j}+h_{i+j},\qquad i,j=1,\ldots, n+2
\end{equation}
with independent (real or complex) coefficients $t_j,\; j=0,1,\ldots n+1$ and real $h_j,\;j=2,\ldots,2n+4$.
The total number of parameters is now $N_t=1+\beta(n+1)+(2n+3)$. 

As in the previous subsections  to find 'zero modes' one has calculate the number of  matrices from this ensemble such 
\begin{equation}
\kT+\kH=\lambda \mathbb{1}.
\end{equation}
This condition now yields the restrictions 
\begin{align}
\kt_0+\kh_{2i}&=\lambda,\qquad i=1,\ldots, n+2,\\
\kt_{i-j}+\kh_{i+j}&=0,\qquad i=j+1,\ldots, n+2 . 
\end{align}
The first condition entails that $\kh_{2i}=\lambda-\kt_0$ for $i=1,\ldots, n+2$, while the second condition, for $1\leq j\leq n+1$ and $i=j+k$, gives
\begin{equation}
\label{ht1}
\kh_{2j+k}=-\kt_k,\qquad k=1,\ldots n+2-j\,.
\end{equation}
In particular $\kh_{2j+1}=-\kt_1$ for $1\leq j\leq n+1$, and thus all $\kh_j$ are fixed. Equation \eqref{ht1} fixes in turn all $\kt_k$ for $k\geq 2$, and constrains the imaginary part of $\kt_1$ to be zero. Only $\kt_0$, $\lambda$, and the real part of $\kt_1$ remain free, providing three 'zero modes'. Therefore one has $q_n=N_t-3=2n+1+\beta (n+1)$. It means that for independent Toeplitz-plus-Hankel matrices $\gamma_n=2n+\beta(n+1)$. This gives
\begin{eqnarray}
\gamma_n&=3n+1\qquad&\textrm{ for }\beta=1\\
\gamma_n&=4n+2\qquad&\textrm{ for }\beta=2.
\end{eqnarray}

\subsection{Hankel matrices}\label{wdensrevH}
Let us now turn to an ensemble of pure Hankel matrices $H_{jk}$, with entries of the form $h_{i+j}$. While in all cases considered so far the identity matrix was a member of the ensemble, this is not the case anymore : a Hankel matrix cannot have all eigenvalues equal, since the identity matrix is not of Hankel form (this also reflects in the peculiar density of eigenvalues for this ensemble as compared with the others, as will be illustrated in Fig.~\ref{density_all} below). Spectra with $n+2$ almost degenerate eigenvalues can be nevertheless obtained by considering matrices of size $N=2n+2$. Indeed, the $N\times N$ matrix 
\begin{equation}
 d_{jk}(N)=\left \{ \begin{array}{cc}1,& j+k\equiv 0\; \mathrm{mod} \;N \\
0, &j+k\not \equiv 0\; \mathrm{mod} \;N \end{array} \right .
\label{H_delta}
\end{equation} 
with $N=2n+2$ has $n+2$ eigenvalues equal to $1$ and $n$ eigenvalues $-1$, therefore in the vicinity of $d(N)$ matrices have $n+2$ almost degenerate eigenvalues close to 1. Therefore Hankel matrices of the form 
\begin{equation}
H_{jk}=\lambda d_{jk}(N)+ \epsilon \kH_{jk}, \qquad N=2n+2\,,
\label{identity_Hankel}
\end{equation}
with $\kH_{jk}=\kh_{j+k}$ and $\epsilon\to 0$, have $n+2$ eigenvalues close to $\lambda$.

 A real symmetric $N\times N$ Hankel matrix $h_{i+j}$ is determined by $N_t=2N-1$ parameters $h_{2},\ldots, h_{2N}$. Following the reasoning of  the previous subsections,  to obtain the small-$s$ behavior of $P_n(s)$ one has  to calculate the number of  independent variables, apart from those that do not affect differences between eigenvalues. For matrices of the form close to the identity matrix as in \eqref{close_to_identity} all eigenvalues are of the order of $\epsilon$ around $\lambda$. 
By contrast, in the case of matrices \eqref{identity_Hankel} there are certain $\kH$ such that deviations of perturbed eigenvalues are of the order of   $\epsilon^2$ around $\lambda$. As we are looking for $P_n(s)$ only at the lowest order on $s$ these 'almost-zero modes' will give higher-order contribution and have to be excluded from the counting of independent variables.   
  
Since $d(N)^2=\mathbb{1}$ is the identity matrix,  the square of matrix \eqref{identity_Hankel} is $H^2=\lambda^2\mathbb{1}+\lambda \epsilon (d(N) \kH+\kH d(N))+\epsilon^2 \kH^2$.  If the anti-commutator of matrices $d(N)$ and $\kH$ is zero,
\begin{equation}
d(N) \kH+\kH d(N)=0\,,
\label{small_hankel}
\end{equation}
then eigenvalues of $H$ are $\pm \lambda +\mathcal{O}(\epsilon^2)$  and such matrices  corresponds to 'almost zero modes', and perturbations $\kH$ in the direction specified by Eq.~\eqref{small_hankel} will not affect eigenvalue spacings (at lowest order).
The requirement  \eqref{small_hankel}  leads to the following conditions 
\begin{eqnarray}
\kh_k+\kh_{2N-k}&=&0,\quad 2\leq k\leq N \nonumber\\
\kh_{N+k}+\kh_{2N-k}&=&0,\quad 1\leq k\leq N.
\label{relations_kh}
\end{eqnarray} 
As a consequence, for $2\leq k\leq N-2$ we must have $\kh_k=\kh_{N+k}=-\kh_{2N-k}=-\kh_{N-k}$ (in particular this fixes $\kh_{N/2}=0$). For $k=N-1$ we must have $\kh_{N-1}=-\kh_{N+1}=\kh_{2N-1}$. For $k=N$ we get $\kh_{N}=\kh_{2N}=0$. The remaining free parameters are thus $\kh_2,\kh_3,...,\kh_{N/2-1}$, $\kh_{N-1}$ and of course $\lambda$. 
This gives in total $n+1$ 'almost zero modes', so that $q_n=N_t-n-1=3n+2$. For random Hankel matrices with independent elements we thus get the gamma distribution \eqref{gamma_pdf} with
\begin{equation}
\gamma_n=3n+1.
\label{gamma_Hankel}
\end{equation}

An alternative, perhaps more transparent way of understanding the origin of condition \eqref{relations_kh} is to notice that for Hankel matrices the condition $\kh_k=\kh_{N+k}$ with $k=2,\ldots, N$, obtained below \eqref{relations_kh}, defines the so-called Hankel circulant matrices, which can easily be diagonalised in Fourier space (cf.~Eq.~\eqref{fourier_hankel}). It is plain that for Hermitian matrices $H_{ij}=h_{i+j}$ of the form \eqref{identity_Hankel} with such a property eigenvalues are given by
\begin{equation}
 \lambda_n=\left \{ \begin{array}{cc}  
 \pm  | \xi_n  |, & n\neq \tfrac{1}{2}N, \;N\\
 \xi_n &n=\tfrac{1}{2}N, \;N\end{array}\right .,\qquad \xi_n=\sum_{r=1}^N \kh_{r+N} \mathrm{e}^{-2\pi \mathrm{i} rn/N} .
  \end{equation}
In the problem considered here one has $h_{N/2}=h_N=\lambda$. Therefore 
\begin{equation}
\xi_n=\lambda+ \sum_{r=1}^{N-1} \kh_{r+N} \mathrm{e}^{-2\pi \mathrm{i} rn/N}.
\end{equation}
In order that the modulus $|\xi_n|$ equals $\lambda+\mathcal{O}(\kh^2)$, it is necessary that the sum in this equation be a pure imaginary for all $n\neq N/2,N$. This requires $\kh_{r+N}+\kh_{2N-r}=0$, which gives back \eqref{relations_kh}. 

The heuristic results of this Section are that for all matrix families given by Eq.~\eqref{main_matrices} the $n$th nearest-neighbor distributions $P_n(s)$ should be well described by the gamma distribution \eqref{gamma_pdf} with $\gamma_n$ summarized in Table~\ref{table}.


\begin{table}
\centering
\begin{tabular}{  | p{ 18em } | c  |}
\hline \hline
Matrix type  & $\gamma_n$  \\
\hline 
Complex Toeplitz matrices & $2n+1$ \\
Special Toeplitz-plus-Hankel matrices & $2n+1$ \\ 
Hankel matrices & $3n+1$\\
Real Toeplitz-plus-Hankel  matrices &  $3n+1$\\
Complex Toeplitz-plus-Hankel matrices & $4n+2$ \\
\hline \hline
\end{tabular} 
\caption{Values of $\gamma_n$ for different types of Hermitian matrices with independent elements.}
\label{table}
\end{table}

\section{Other spectral properties}\label{other_spectral_properties}

\subsection{Level compressibility}\label{compressibility}
A characteristic property of models with intermediate statistics is the non-trivial value of the level compressibility $\chi$, which is determined through the limiting behavior of the variance of the number of eigenvalues (normalized to unit density) in an interval of length $L$. If $N(L)$ is the number of eigenvalues inside the interval $L$ then
\begin{equation}
\Sigma^2(L)\equiv \langle (N(L)-L)^2 \rangle \underset{L\to \infty}{\sim} \chi L\,,
\label{sigma_2}
\end{equation}
where  the average is taken over different realizations of random matrices. For the usual Wigner-Dyson random matrix ensembles $\chi=0$ and for the Poisson distribution $\chi=1$. For intermediate statistics it is assumed \cite{shklovskii, altshuler} that
\begin{equation}
0<\chi<1.
\end{equation}
The calculation of the number variance requires the knowledge of the two-point correlation function 
$R_2(s)$, determined as the probability that two eigenvalues are separated by a distance $s$. Since there is an arbitrary number of eigenvalues inside this interval, it is plain that it equals the sum over all nearest-neighbor distributions
\begin{equation}
R_2(s)=\sum_{n=0}^{\infty} P_n(s). 
\label{R2}
\end{equation} 
The gamma distributions proposed above for $P_n(s)$ are only approximations to unknown expressions  and small errors hardly visible in the nearest-neighbor distributions may lead to considerable deviations in the infinite sum for $R_2(s)$. Nevertheless, it is instructive to see what they give for the compressibility. 

The distribution $P_n(s)$ has the form $s^{\gamma_n} \exp (-\gamma_n s/(n+1))$, where $\gamma_n=pn+k$ with $p$ and $k$ independent of $n$. It has is maximum at $s=n+1$ (which coincides with its mean value $s=n+1$ given by the normalization \eqref{normalization_ps}). A second-order expansion near this maximum gives a Gaussian with mean value $n+1$ and variance $(n+1)^2/\gamma_n$. At large $n$ the distribution can then be approximated asymptotically by
\begin{equation}
P_n(s)=\frac{1}{\sqrt{2\pi \Sigma^2(n)}}\exp\left ( -\frac{(s-n-1)^2}{2\Sigma^2(n)}\right ), \qquad \Sigma^2(n)=\frac{n}{p}. 
\end{equation} 
This formula can be reversed to get the behavior of $P_n(s)$ as function of $n$ at large fixed $s$ 
 \begin{equation}
P_n(s)=\frac{1}{\sqrt{2\pi \Sigma^2(s)}}\exp\left ( -\frac{(n-s)^2}{2\Sigma^2(s)}\right ), \qquad \Sigma^2(s)=\frac{s}{p}. 
\end{equation} 
Therefore from the definition \eqref{sigma_2} it follows that
\begin{equation}
\Sigma^2(L)\approx \int (n-L)^2P_n(L)\mathrm{d}n=\frac{L}{p}\,,
\end{equation}
 which means that for a gamma distribution with $\gamma_n=pn+k$ 
 \begin{equation}
\label{chi1p}
 \chi=\frac{1}{p}. 
 \end{equation}
 In other words, if $P_n(s)$ at large $n$ has exponential decrease as $\exp( -p s)$ then $\chi=1/p$. Such a relation is also valid for all short-range plasma models discussed in \cite{gerland} (cf Appendix~\ref{plasma}).

\subsection{Form factor}\label{formfactor}
The compressibility can be recovered alternatively from the asymptotic behavior of the two-point correlation form factor as 
\begin{equation}
\chi=\lim_{\tau\to 0} K(\tau),
\label{chi_from_K}
\end{equation}
where the form factor is the Fourier transform of the spectral two-point correlation function, defined as
\begin{equation}
\label{defktau}
K(\tau)=\int_{-\infty}^{\infty} R_2(s)\mathrm{e}^{2\pi \mathrm{i} \tau s}\mathrm{d}s. 
\end{equation} 
The Fourier transform \eqref{defktau} can be evaluated by introducing the Laplace transform of the sum \eqref{R2}, which yields
\begin{equation}
K(\tau)=1+2 \mathrm{Re}\, g(2\pi \mathrm{i} \tau)\,,
\label{K_from_g}
\end{equation} 
where the function $g$ is defined as
\begin{equation}
\label{gtsum}
g(t)= \sum_{n=0}^{\infty} g_n(t),
\end{equation}
with $g_n(t)$ the Laplace transform of $P_n(s)$. Assuming that $\gamma_n=pn+k$ with $p$ and $k$ independent on $n$ one gets from \eqref{gamma_pdf} that
\begin{equation}
g_n(t)=\int_0^{\infty} P_n(s)\mathrm{e}^{- t s}\mathrm{d}s=\left (1+\frac{(n+1)t}{pn+k+1}\right)^{-pn-k-1}.
\end{equation} 
We can then reexpress $g(t)$ as the sum of two terms, 
\begin{equation}
g(t)=f(t) +\exp\left (-\frac{t (p-k-1)}{p+t}\right ) \sum_{n=0}^{\infty}  \left (1+\frac{t}{p} \right )^{-pn-k-1},
\label{sum}
\end{equation}
where
\begin{equation}
f(t)=\sum_{n=0}^{\infty} \left[ \left (1+\frac{(n+1)t}{pn+k+1}\right)^{-pn-k-1}
-\left (1+\frac{t}{p} \right )^{-pn-k-1} \exp\left (-\frac{t (p-k-1)}{p+t}\right ) \right ]\,.
\label{f_finite}
\end{equation}
The function $f$ has a finite limit at $t\to 0$ and verifies $f(0)=0$.
The second term in \eqref{sum} is easily calculated and yields
\begin{equation}
g(t)=f(t)+\frac{(1+\frac{t}{p})^{p-k-1}}{(1+\frac{t}{p})^p-1}\exp\left(-\frac{t (p-k-1)}{p+t}\right).
\end{equation}
While \eqref{gtsum} diverges when evaluated numerically, the form \eqref{sum}--\eqref{f_finite} allows to obtain a theoretical prediction for the form factor, using Eq.~\eqref{K_from_g}. This expression will be compared with numerical computations in the next section.

From the second term it follows that
\begin{equation}
g(t)\underset{t\to 0}{\sim} \frac{1}{t}-\frac{p-1}{2p}+\mathcal{O}(t).
\end{equation}
The behavior of $g(t)$ at small $t$ yields, using \eqref{K_from_g},
$\chi \equiv  K(0)=1/p$,
which coincides with the expression in Eq.~\eqref{chi1p}.

\section{Comparison with numerical calculations}\label{numerical_correlation_functions}
We now turn to the numerical determination of the different correlation functions for the families of random matrices considered above. The ensembles are constructed by taking all independent real matrix elements of these matrices (or all real and imaginary parts for complex entries) as Gaussian random variables with zero mean and unit variance. The spectra were obtained by diagonalization of matrices with dimension $N=2^{10}$, and for each family $20\, 000$ realizations of random matrices were taken.

\subsection{Mean density of eigenvalues}
\label{meandensity}
The most basic quantity that characterises the spectrum of a matrix of size $N$ is the density of eigenvalues, defined as
\begin{equation}
\bar{\rho}(E)=\frac{1}{N}\left \langle \sum_{j=1}^N \delta(E-E_j)\right \rangle,
\end{equation}
where the average is taken over different realisations of random parameters. 

\begin{figure}
\begin{center}
\includegraphics[width=.6\linewidth]{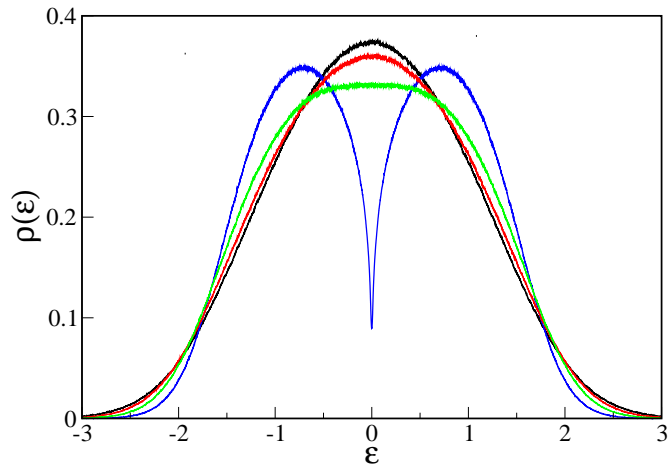}
\end{center}
\caption{Mean normalized densities $\bar{\rho}(\varepsilon)$ for different classes of structured matrices: real and complex Toeplitz (black), independent real Toeplitz-plus-Hankel (red), independent complex Toeplitz-plus-Hankel (green), and Hankel (blue).}
\label{density_all}
\end{figure}

It is well-known that for the usual Wigner-Dyson random matrix ensembles the mean density of states follows the Wigner semi-circle law. On the other hand, for real symmetric Toeplitz and Hankel matrices, it was shown in \cite{hammond, bryc} that the mean densities (rescaled by $\sqrt{N}$) converge when $N\to\infty$ to  nontrivial symmetric distributions  depending only on the variance of matrix elements. To meaningfully compare mean densities of different matrix ensembles one should therefore rescale them in such a way that
\begin{equation}
\int E^2 \bar{\rho}(E)\mathrm{d}E=1.
\end{equation}
This can be achieved by rescaling energy levels as $\varepsilon=E/\sigma$, with  $\sigma^2=\langle \mathrm{Tr}(M^2) \rangle/N$. 
For matrices \eqref{main_matrices} with independent entries of zero mean and unit variance, one readily gets that for large $N$ the rescaling factor is $\sigma^2=N$ for real symmetric Toeplitz matrices and Hankel matrices, $\sigma^2=2N$ for complex Toeplitz matrices and real Toeplitz-plus-Hankel matrices with independent entries, and $\sigma^2=3N$ for complex Toeplitz-plus-Hankel matrices. The rescaled mean densities are presented in Fig.~\ref{density_all}. 

\subsection{Nearest-neighbor distributions} 

In order to compare numerical data with universal analytic predictions, we need to perform what is known as an 'unfolding' procedure. Assuming that  eigenvalues are ordered, $E_{j+1}\geq E_{j}$, unfolded eigenvalues $e_j$ are defined as
\begin{equation}
e_j=\bar{N}(E_j),
\end{equation} 
where $\bar{N}(E)$ is the cumulative mean density
\begin{equation}
\bar{N}(E)=\int_{-\infty}^E\bar{\rho}(E^{\prime})\mathrm{d} E^{\prime}. 
\end{equation}
The unfolded eigenvalues have unit mean density. Nearest-neighbor distributions $P_n(s)$ are then calculated from a small interval of the unfolded spectrum around the maximum of $\bar{\rho}(E)$. In practice we took an interval containing $1/4$ of the total number of levels around the center of the spectrum. For Hankel matrices however, because of the unusual two-peak form of the density (see Fig.~\ref{density_all}), eigenvalues were taken around the right peak only.

\begin{figure}
 \begin{center}
 \includegraphics[width=.59\linewidth]{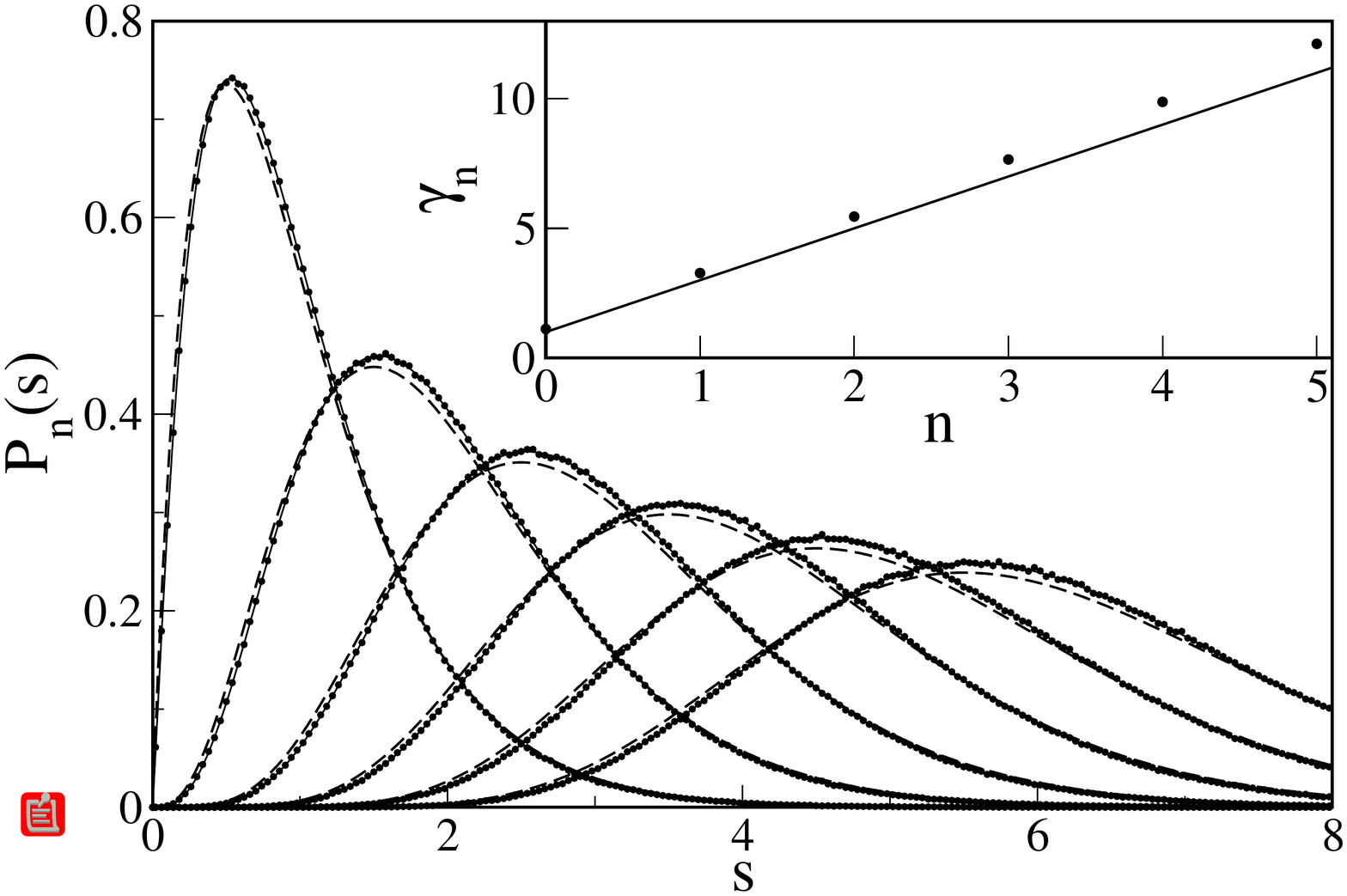}\\
\includegraphics[width=.59\linewidth]{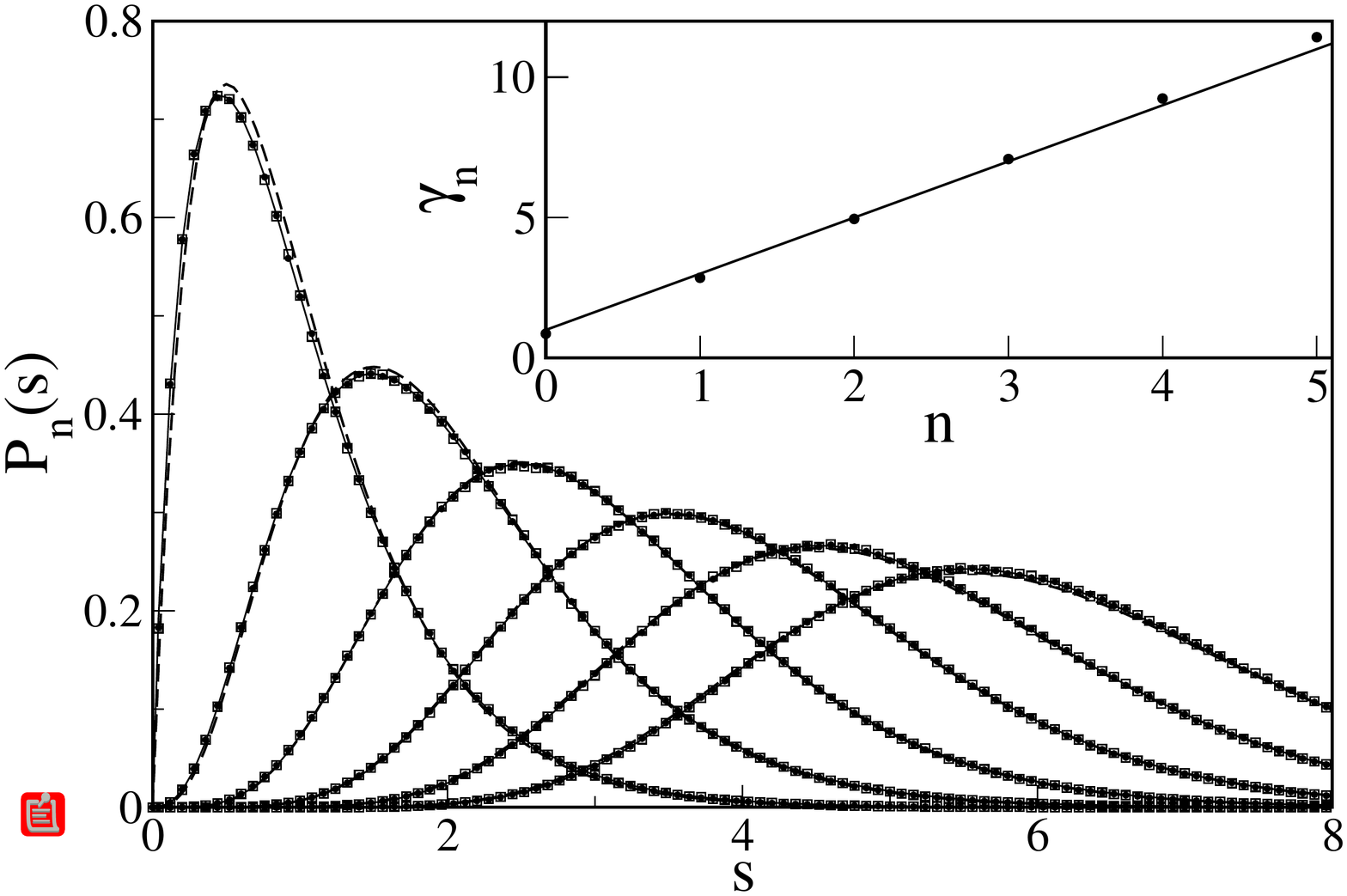}
\end{center}
\caption{
 $P_n(s)$ for $0\leq n\leq 5$ for (top) complex Toeplitz matrices  (open black squares) and (bottom) special $T+H$ matrices (filled black circles) and special $T-H$ matrices (open black squares) given by \eqref{even}. Black dashed lines are gamma distributions \eqref{gamma_pdf} with $\gamma_n=2n+1$ (semi-Poisson distribution). Black solid lines are gamma distributions fitted with a single fitting parameter $\gamma_n$. The corresponding fitted values of $\gamma_n$ are given in Table \ref{table_fit} and plotted in the insets (filled black circles) together with the semi-Poisson prediction $\gamma_n=2n+1$ (solid black line).\label{ps_T}}
\end{figure}

\begin{figure}
\begin{center}
\includegraphics[width=.59\linewidth]{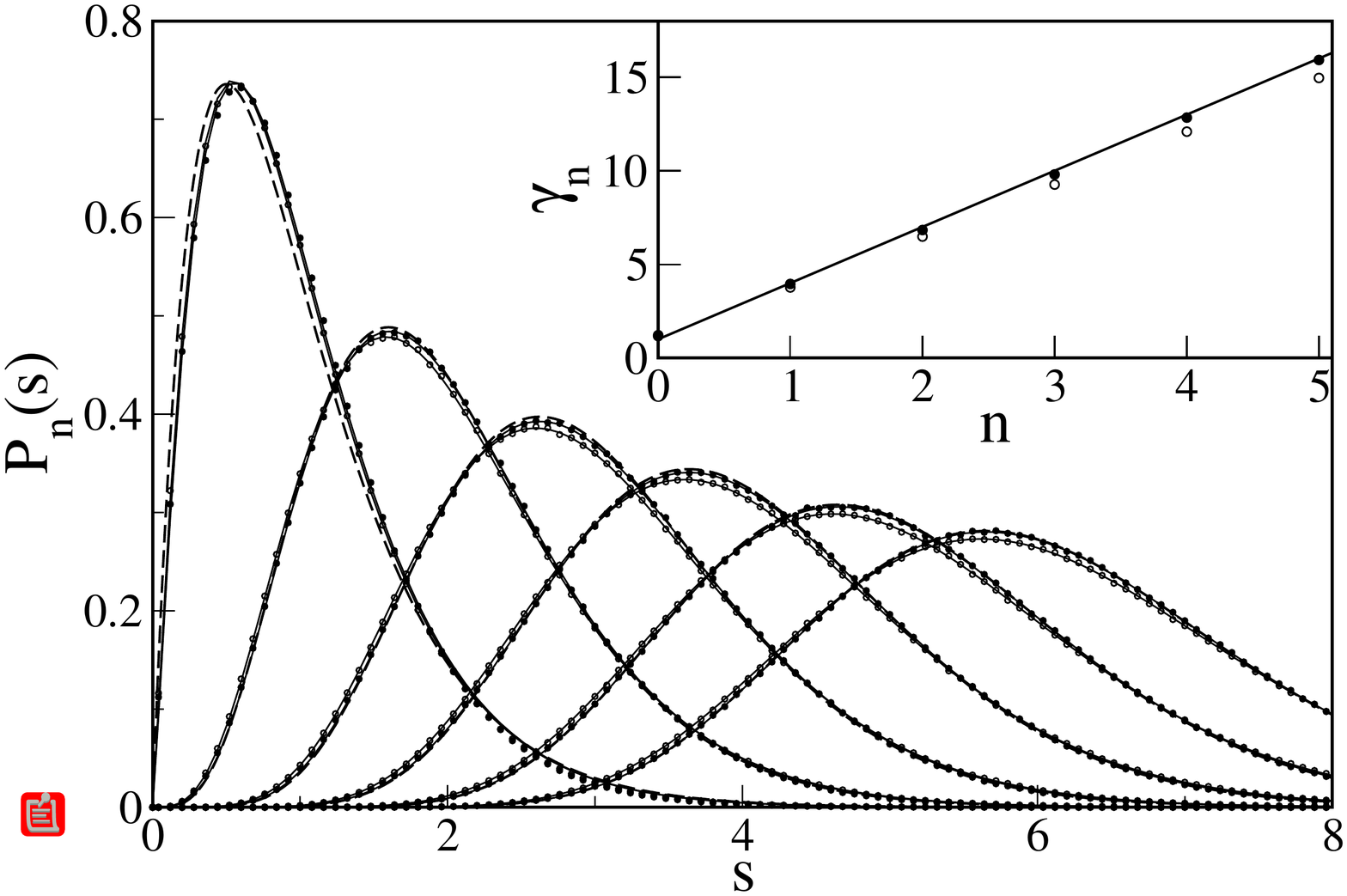}\\
\includegraphics[width=.59\linewidth]{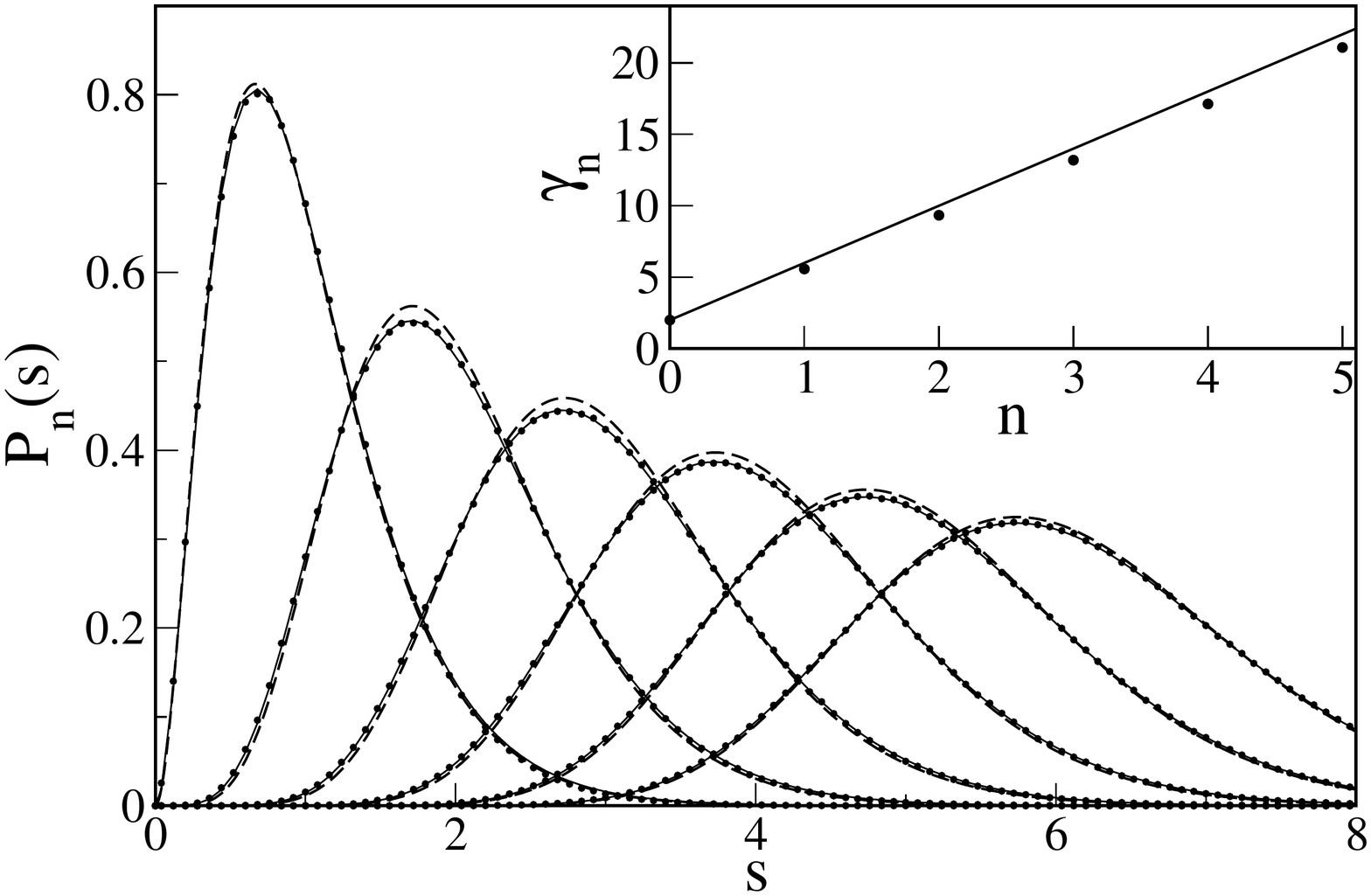}
\end{center}
\caption{
(Top) $P_n(s)$ for $0\leq n\leq 5$ for Hankel matrices  (open black  circles) and real Toeplitz-plus-Hankel matrices  (filled black circles). Black dashed  lines are the gamma distributions \eqref{gamma_pdf} with $\gamma_n=3n+1$. The inset displays the values of $\gamma_n$ obtained by a one-parameter fit (same symbols as in the main panel) and given in Table \ref{table_fit}, black solid line is the prediction $\gamma_n=3n+1$. (Bottom) Same for complex Toeplitz-plus-Hankel matrices (filled black circles). Black dashed lines are the gamma distributions \eqref{gamma_pdf} with $\gamma_n=4n+2$, black solid lines are the gamma distributions  obtained by a one-parameter fit. The inset shows the fitted values of $\gamma_n$ (filled black circles) and the prediction $\gamma_n=4n+2$.}
\label{ps_H_T+H}
\end{figure}

The results for nearest-neighbor distributions $P_n(s)$ with $0\leq n\leq 5$ are presented in Fig.~\ref{ps_T}
for complex Toeplitz matrices and for special $T\pm H$ matrices, together with the theoretical gamma distributions \eqref{gamma_pdf}. The same quantities for Hankel matrices and for real and complex independent Toeplitz-plus-Hankel matrices are plotted in Fig.~\ref{ps_H_T+H}. The theoretical distributions \eqref{gamma_pdf} agree quite well with numerical calculations, despite the fact that these formulas have no free parameters. Note that numerically calculated spectral correlation functions of Hankel and real Toeplitz-plus-Hankel matrices are close to each other (see Fig.~\ref{ps_H_T+H} top), although their spectral densities, plotted in Fig.~\ref{density_all}, are very different.  

In order to further improve the formulas for $P_n(s)$, one can replace the term $s^{\gamma}$ in Eq.~\eqref{gamma_pdf} by a polynomial is $s$, as happens in the short-range plasma model \cite{gerland}. This is briefly discussed in Appendix~\ref{plasma}. A drawback of such an approach is that normalization of the expressions lead to quite cumbersome formulas. We found that the simplest way to improve our surmise for $P_n(s)$ is to use the same gamma distribution \eqref{gamma_pdf} with normalization \eqref{correct_norm}, but with a value of $\gamma_n$ obtained from a one-parameter fit of the data (the only free parameter being $\gamma_n$). Surprisingly, such ad hoc fits  work very well. The results are displayed at Figs.~\ref{ps_T} and \ref{ps_H_T+H}, and the fitting curves entirely go through numerical points. In the insets of these figures we compare the fitted values of $\gamma_n$, given in Table~\ref{table_fit}, with the predictions of Table~\ref{table}.

 \begin{table}
\centering
\begin{tabular}{ | p{18em} | c | c | c | c | c | c |}
\hline \hline
Matrix type   & $\gamma_0$ &  $\gamma_1$ &$\gamma_2$ &$\gamma_3$ &$\gamma_4$& $\gamma_5$  \\
\hline 
Complex Toeplitz matrices & $ 1.12$& $3.28$ & $5.45$ & $ 7.66 $& $9.88$& $12.12$\\
Special Toeplitz-plus-Hankel matrices & $0.86 $ & $2.86 $& $4.95 $& $7.08 $& $9.24 $& $11.42$ \\ 
Hankel matrices & $1.17$ & $3.77 $ & $6.48 $& $9.27 $ & $12.09 $& $14.96$ \\
Real Toeplitz-plus-Hankel  matrices &  $1.22 $ & $3.96 $ & $6.83 $ & $9.81 $ & $12.84 $ & 15.92  \\
Complex Toeplitz-plus-Hankel matrices & $2.00 $ & $5.58 $ & $9.33 $& $13.20 $& $17.13$ & $21.08 $ \\
\hline \hline
\end{tabular} 
\caption{Values of fitted $\gamma_n$ with $n=0,\ldots,5$ for different  matrices}
\label{table_fit}
\end{table}

\subsection{Two-point correlation form factor}

\begin{figure}
\begin{minipage}{.49\linewidth}
\begin{center}
\includegraphics[width=.99\linewidth]{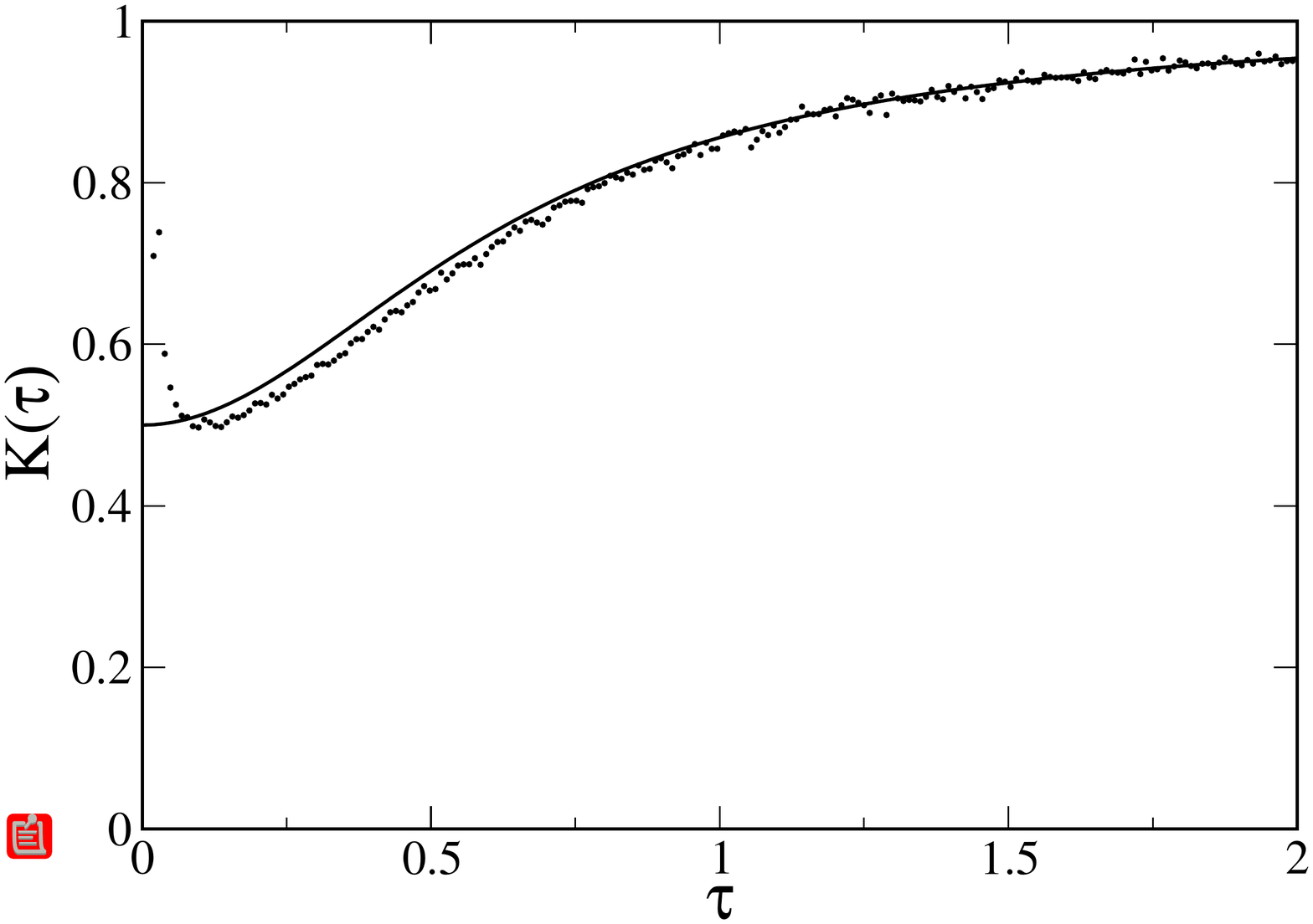}\\
(a)
\end{center}
\end{minipage}
\begin{minipage}{.49\linewidth}
\begin{center}
\includegraphics[width=.99\linewidth]{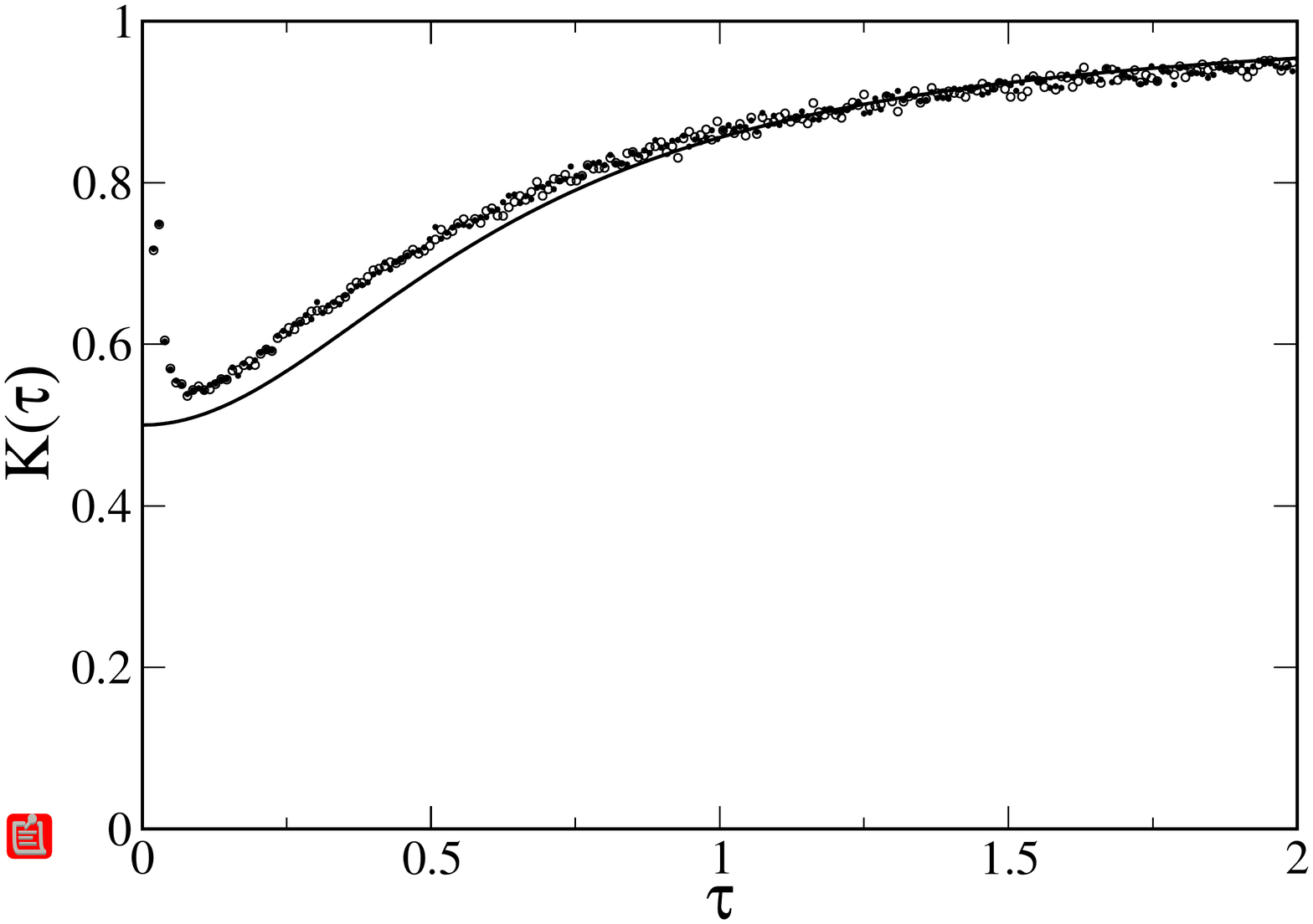}\\
(b)
\end{center}
\end{minipage}
\caption{
(a) The form factor of (a) complex Toeplitz  matrices (filled black circles), (b) special $T+H$ matrices (filled black circles) and special $T-H$ matrices (open black circles).
Circles are the numerical computations using \eqref{K_definition}, solid black line is the form factor of the gamma distribution with $\gamma_n=2n+1$ (i.e.~the semi-Poisson distribution) calculated from \eqref{sum} and \eqref{f_finite}.} 
\label{fig_formfactor_Toeplitz}
\end{figure}

\begin{figure}
\begin{minipage}{.49\linewidth}
\begin{center}
\includegraphics[width=.99\linewidth]{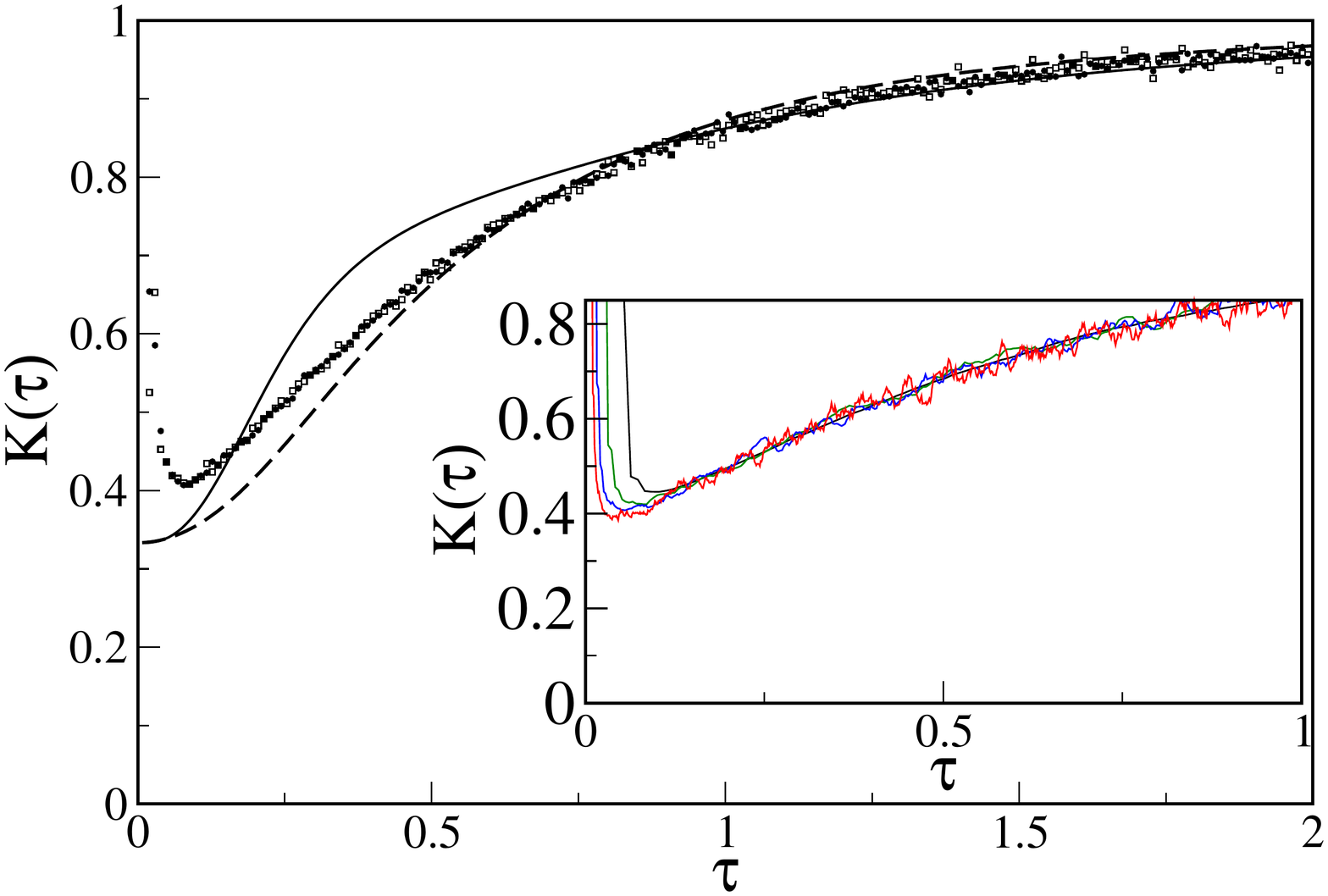}\\
(a)
\end{center}
\end{minipage}
\begin{minipage}{.49\linewidth}
\begin{center}
\includegraphics[width=.99\linewidth]{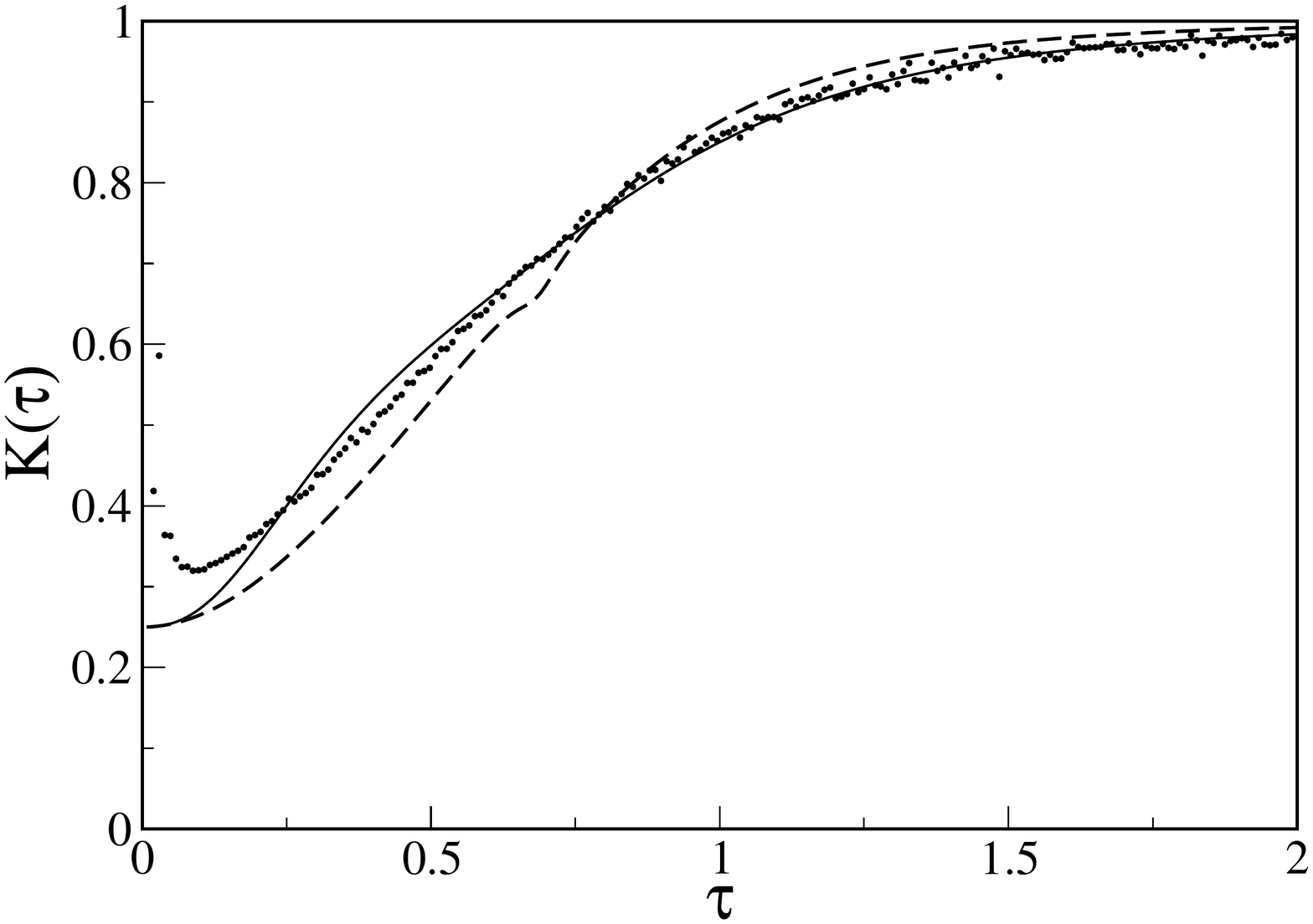}\\
(b)
\end{center}
\end{minipage}
\caption{
(a) The form factor of Hankel matrices (filled black circles) and independent real Toeplitz-plus-Hankel matrices (open black squares). Inset: evolution of the form factor near the origin  for Hankel matrices with increasing matrix dimension $N=2^n$ with  $n=10$ (black), $n=11$ (green), $n=12$ (blue), and $n=13$ (red). 
(b) The form factor of independent complex Toeplitz-plus-Hankel matrices (filled black circles).
Solid black line is the form factor of gamma distribution with:  (a) $\gamma_n=3n+1$, and (b) $\gamma_n=4n+2$ calculated from \eqref{sum} and \eqref{f_finite}. Dashed line is the form factor for short-range plasma model with the same value of $\gamma_n$, given by \eqref{g_first} and \eqref{g_second}.} 
\label{fig_formfactor}
\end{figure}

 The two-point correlation form factor $K(\tau)$ is the Fourier transform of the spectral two-point correlation function $R_2(s)$. It can be expressed in terms of the unfolded spectrum as
\begin{equation}
K(\tau)=\frac{1}{N} \left \langle  \left | \sum_{j=1}^N\mathrm{e}^{2\pi \mathrm{i} e_j \tau }\right |^2 \right \rangle 
\label{K_definition}
\end{equation}
(see e.g.~\cite{Haa91}). The form factor of complex Toeplitz matrices and special $T\pm H$ matrices \eqref{even}, computed numerically from \eqref{K_definition}, is displayed in Fig.~\ref{fig_formfactor_Toeplitz}. In Fig.~\ref{fig_formfactor}a we plot the same function for Hankel and real independent Toeplitz-plus-Hankel matrices, and in Fig.~\ref{fig_formfactor}b for complex independent Toeplitz-plus-Hankel matrices. The form factors for $\gamma_n=2n+1, 3n+1, 4n+2$, calculated from the above formula \eqref{K_from_g} are presented in Figs.~\ref{fig_formfactor_Toeplitz} and \ref{fig_formfactor} together with the exact form factors of corresponding short-range plasma models presented in Appendix~\ref{plasma}. 

It is known that numerical calculation of the spectral compressibility from finite-dimensional matrices is subtle.  The point is that the compressibility is defined  either from the large-$L$  behavior of the number variance  \eqref{sigma_2} or from the limiting value of the form factor at small argument \eqref{chi_from_K}. In such definitions it is implicitly assumed that the limit $N\to\infty$ is taken first, but in  numerics one fixes the matrix dimension. This inevitable inversion of the limits is responsible for the sudden increase of $K(\tau)$ evident in the above figures. The same phenomenon is clearly seen even in \eqref{K_definition}, where formally $K(0)=N\underset{N\to \infty}{\longrightarrow}\infty$. More and more realizations of random parameters are needed to get correctly the limiting value $K(0)$. This is illustrated in the inset of Fig.~\ref{fig_formfactor}.

\subsection{Fractal dimensions}\label{secfractaldim}
Previous sections have shown that spectral statistics of Toeplitz and Hankel matrices, or their sums, are of intermediate type.
 Such a behavior of spectral fluctuations has been associated with multifractal properties of eigenstates \cite{Weg80, Aok83, CasPel86}. Multifractals are objects that display fluctuations at all scales and are characterized by the existence of a whole range of fractal dimensions. 

Contrary to eigenvalues, eigenvectors (and thus multifractality and fractal dimensions) depend on the chosen basis. For models with intermediate-type statistics eigenfunctions are typically fully extended in coordinate space but multifractal in Fourier space. We thus introduce eigenfunctions $\hat{\Psi}_p(E)$, defined as the Fourier transform of eigenvectors $\Psi_j(E)$ through the usual expression
\begin{equation}
\label{psipsihat}
\hat{\Psi}_p(E)=\frac{1}{\sqrt{N}} \sum_{j=1}^N \mathrm{e}^{2\pi \mathrm{i} jp/N}\Psi_j(E)\,,
\end{equation}
where $\Psi_j(E)$ is the eigenvector corresponding to the eigenvalue $E$ and it is assumed that this vector is normalized as $\sum_{j=1}^N|\Psi_j(E)|^2=1$. These functions could be also calculated by the diagonalization of Fourier transform of matrices \eqref{main_matrices}. The Fourier transform of the  Toeplitz matrix is given in \cite{bogomolny}. For Hankel matrices one gets
\begin{eqnarray}
\hat{H}_{mn}&=&\frac{1}{N} \sum_{j,k=1}^N h_{j+k}\mathrm{e}^{2\pi \mathrm{i} (km-jn)/N}\nonumber\\
&=&\xi_n d_{mn}(N) -
(1-d_{mn}(N)) \left [\frac{\eta_n}{1-\mathrm{e}^{-2\pi \mathrm{i} (m+n)/N}}+ \frac{\eta_m^*}{1-\mathrm{e}^{2\pi \mathrm{i} (m+n)/N}}\right ]. 
\end{eqnarray}  
Here matrix $d_{mn}(N)$ was defined in \eqref{H_delta}  and 
\begin{eqnarray}
\xi_n&=&\frac{1}{N}\sum_{r=2}^N (h_r-h_{r+N})(r-1)\mathrm{e}^{-2\pi \mathrm{i} rn/N}+ \sum_{r=1}^N h_{r+N} \mathrm{e}^{-2\pi \mathrm{i} rn/N},\nonumber\\
\eta_n&=&\frac{1}{N}\sum_{r=2}^N(h_r-h_{r+N})\mathrm{e}^{-2\pi \mathrm{i} rn/N}\,.
\label{fourier_hankel}
\end{eqnarray}

In order to define multifractal dimensions one calculates the moments of eigenvectors and their scaling with matrix dimension (see, e.g., \cite{mirlin} and references therein)
\begin{equation}
\left \langle \sum_{p=1}^N |\hat{\Psi}_p(E)|^{2q}\right \rangle \underset{N\to\infty}{\sim} C\,N^{-\tau(q)}\,.
\label{moments}
\end{equation} 
The average in the above expression is taken over different realizations of random parameters and over all eigenvalues in a small energy window around $E$. 
The exponent $\tau(q)$ determines the dependence of the $q$th moment with $N$. If an eigenvector 
has a small number of large components then $\tau(q)=0$. If all components are of comparable magnitudes then, from normalization, $|\Psi_j(E)|^2\sim N^{-1}$, and thus $\tau(q)=q-1$. The ratios
\begin{equation}
D_q=\frac{\tau(q)}{q-1}
\end{equation}
are called (multi)fractal dimensions and they are the main characteristics of statistical properties of eigenvectors.  
For localized states $D_q=0$ and for fully extended states $D_q=1$. Systems for which $D_q$ differs from these extreme values and depend on $q$ are called multifractal. 

\begin{figure}
\begin{minipage}{.49\linewidth}
\begin{center}
\includegraphics[width=.99\linewidth]{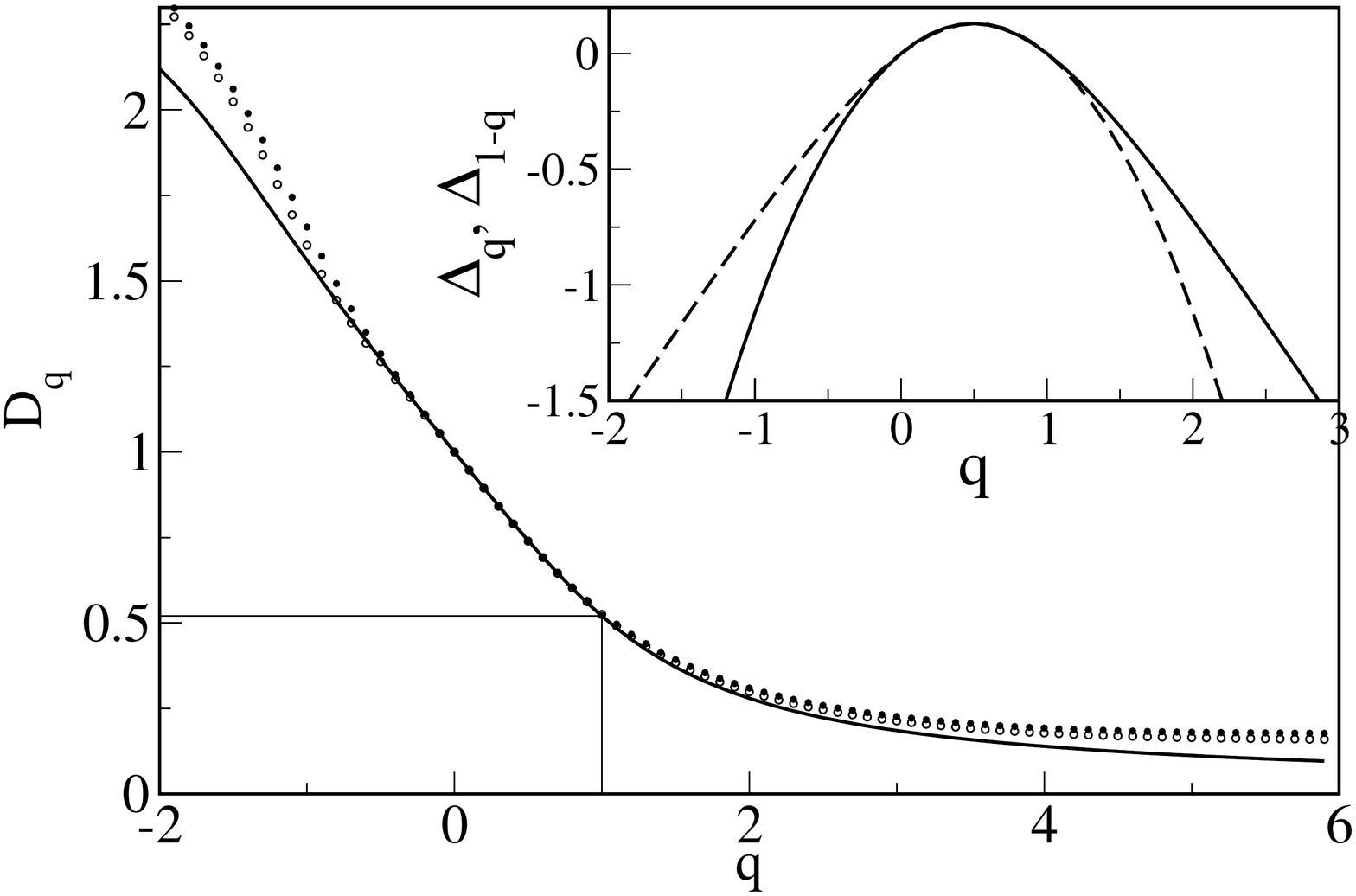}\\
(a)
\end{center}
\end{minipage}
\begin{minipage}{.49\linewidth}
\begin{center}
\includegraphics[width=.99\linewidth]{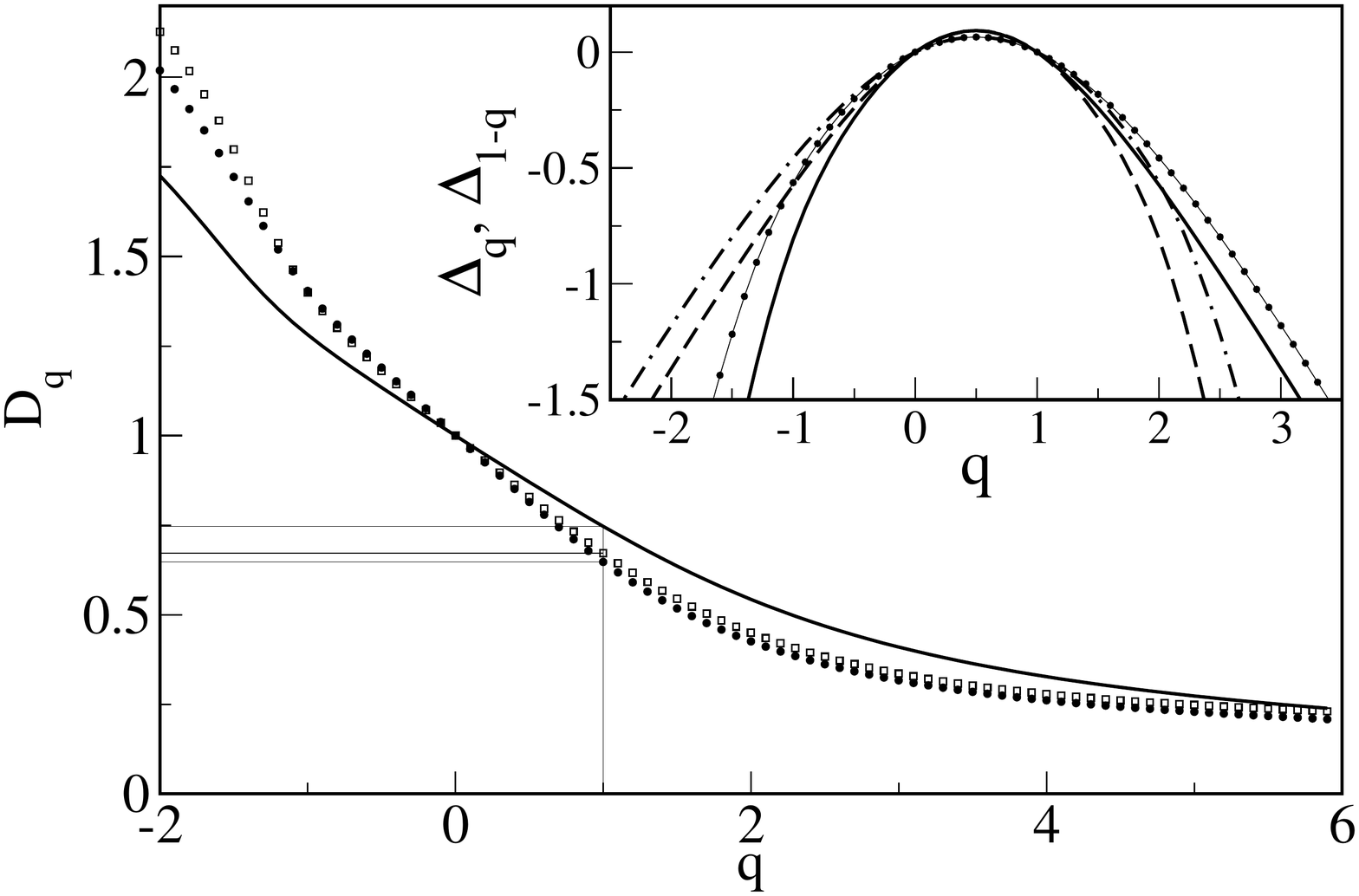}\\
(b)
\end{center}
\end{minipage}
\caption{Fractal dimensions in the Fourier space for (a)  complex Toeplitz matrices (solid black line) and for special $T+H$ matrices  (filled black circles) and $T-H$ matrices (open black circles) given by \eqref{even} and (b) Hankel matrices (filled black circles), independent  real Toeplitz-plus-Hankel matrices (open black squares), and independent complex Toeplitz-plus-Hankel matrices (solid black line). Points corresponding to $D_1$ are indicated by thin straight lines. In the insets  $\Delta_q$ and $\Delta_{1-q}$ are plotted together for (a) complex Toeplitz matrices 
(solid line is $\Delta_q$ and dashed line is $\Delta_{1-q}$) and for (b) Hankel matrices 
($\Delta_q$ is indicated by solid line and $\Delta_{1-q}$ by dashed line), and for independent complex Toeplitz-plus-Hankel matrices ($\Delta_q$ is indicated by filled black circles connected by a thin line 
and $\Delta_{1-q}$ by dashed-dotted line).}
\label{fractal_dim_H_T+H}
\end{figure}

For the ensembles of random matrices considered here, eigenvectors $\Psi_j(E)$ were obtained by exact diagonalization and Fourier transformed according to \eqref{psipsihat}. The exponents $\tau(q)$ were extracted from a linear fit of the logarithm of moments \eqref{moments} as a function of $\ln N$, for data from $N=2^{7}$ ($8000$ realizations) to $N=2^{12}$ ($200$ realizations) in a small window of eigenvalues around the maximum of the density. The numerical results are displayed in Fig.~\ref{fractal_dim_H_T+H}. A first observation is that for all the matrix ensembles considered here, fractal dimensions are non-trivial (i.e., different from $0$ and $1$). As was the case for the spectra, fractal dimensions of Hankel and independent real Toeplitz-plus-Hankel matrices are very close to each other. Though for $q<-1$ they seem to deviate, big numerical uncertainties in this region, due to small wavefunction values taken at a negative power, do not permit to get a clear-cut conclusion. 

There are practically no general  results for fractal dimensions. In Ref.~\cite{delta}, based on the nonlinear sigma model,  it was conjectured that the anomalous dimensions defined by
\begin{equation}
\Delta_q=(D_q-1)(q-1)
\end{equation}
should satisfy the following symmetry relation
\begin{equation}
\Delta_q=\Delta_{1-q}\,.
\label{symmetry}
\end{equation}
In the inset of Fig.~\ref{fractal_dim_H_T+H}, values of $\Delta_q$ and $\Delta_{1-q}$ are plotted for Toeplitz, Hankel,  and complex Toeplitz-plus-Hankel matrices. It is clear that the relation \eqref{symmetry} is valid only in the interval $|q|<1$ where fractal dimensions are practically linear. At larger values of $q$ this relation numerically breaks down.

The quantity  $D_1$ has a special importance as it is a kind of eigenfunction entropy
\begin{equation}
\sum_{p=1}^N |\hat{\Psi}_p|^2 \ln|\hat{\Psi}_p|^2\underset{N\to\infty}{\sim} -D_1 \ln N.
\end{equation} 
In Ref.~\cite{BogGirConj} it was conjectured that 
\begin{equation}
\label{conjecture}
D_1+\chi=1\,,
\end{equation}
where $\chi$ is the level compressibility discussed in Section~\ref{compressibility}. Numerically one finds that 
 for complex Toeplitz matrices $D_1\approx 0.52$, for special $T\pm H$ matrices $D_1\approx 0.52$, for Hankel matrices $D_1\approx 0.65$, for real Toeplitz-plus-Hankel matrices $D_1\approx 0.67$, and for complex Toeplitz-plus-Hankel matrices $D_1\approx 0.75$. These numerical values are quite close to the theoretical values $1/2$, $2/3$ and $3/4$ expected from Eqs.~\eqref{chi1p} and \eqref{conjecture}, which indicates that the conjecture \eqref{conjecture} (approximatively) extends to the all the above random matrix ensembles.

\section{Summary}\label{conclusion} 

Toeplitz, Hankel, and Toeplitz-plus-Hankel matrices are probably the oldest and the best investigated classes of matrices. Though it was known for a long time that these matrices are examples of low-complexity matrices, the investigation of statistical properties of such matrices with very irregular (which may and will  be naturally substituted by random)  elements did not attract wide attention. 

In the present paper such Hermitian matrices with independent and identically distributed random elements were investigated  in detail.  The main spectral correlation functions were calculated using large matrix dimensions and large number of realizations. Special attention was given to the careful determination of  nearest-neighbor distributions of these matrices.  It is demonstrated that these families of matrices display level repulsion at small distances, as for usual Wigner-Dyson ensembles  of random matrices,  but  their nearest-neighbor distributions decrease  exponentially at large argument, contrary to the Wigner-Dyson ensembles where they have Gaussian tails. Combination of these two properties is the characteristic feature of the so-called intermediate spectral statistics observed till now  only in rare special systems, such as the Anderson model at the point of metal-insulator transition, certain pseudo-integrable billiards, intermediate kicked quantum maps, and matrix ensembles related with Lax matrices of integrable models. 

The form factor (the Fourier transform of the two-point spectral correlation function)  is a key characteristics of spectral statistics, and its value at 0, the level compressibility, provides a signature of the nature of these statistics: it takes the value $0$ for the Wigner-Dyson ensembles, $1$ for the Poisson statistics, and an intermediate value for intermediate statistics.  For all matrices considered here the form factors were calculated numerically, and it was demonstrated that they are different from standard random matrix ensembles. In particular, the data indicate that the spectral compressibility of Toeplitz, Hankel, and Toeplitz-plus-Hankel matrices is non-trivial, which is another characteristic feature common to all known intermediate-type statistics. 
Moreover, numerical calculations of statistical properties of the corresponding eigenvectors show that their fractal dimensions are also non-trivial (i.e., different from $0$ or $1$) for all considered matrices, which shows that such eigenfunctions are multifractal. 

Exact analytic results for such statistical distributions are non-existing. Here we developed a simple heuristic method to get approximate formulas  of the nearest-neighbor distributions  for the considered classes of random matrices. They correspond to the normalized gamma distribution with parameter $\gamma_n$ given in Table~\ref{table}.  The obtained expressions have no free parameter and approximate well the numerical results. Adding a single fitting parameter (the parameter $\gamma$ of the gamma distribution) gives almost perfect agreement with the data. Therefore, in the same way as the celebrated Wigner surmise, and because of their simplicity, our formulas can serve as approximations to unknown distributions. We also observed that numerically calculated  correlation functions for Hankel random matrices and independent real Toeplitz-plus-Hankel matrices are surprisingly  close to each other, though their spectral densities are very different.  

The fact that all low-complexity matrices discussed in the paper have this type of statistics clearly shows that intermediate statistics are more widely spread than was considered before and opens new perspectives in random matrix theory. As this type of matrices pervades all branches of physical and mathematical sciences, it is an important challenge to derive analytically correlation functions in these models.

\appendix

\section{Displacement structure for Toeplitz, Hankel, and Toeplitz-plus-Hankel matrices} \label{displacement} 

Let $Z$ be the $n\times n$ matrix 
\begin{equation}
Z_{jk}=\delta_{j-k-1}\,.
 \label{shift_matrix}
\end{equation}
This matrix shifts any matrix $M$ as
\begin{eqnarray}
(Z\,M)_{ij}&=& M_{i-1,j}\Theta(i-1) \longrightarrow  \mathrm{shift\; down}\nonumber\\
(Z^{T}\,M)_{ij} &=& M_{i+1,j}\Theta(n-i) \longrightarrow \mathrm{shift\; up}\nonumber\\
(M\,Z)_{ij}&=&M_{i,j+1} \Theta(n-j) \longrightarrow \mathrm{shift\; left}\nonumber\\
(M\, Z^{T})_{ij}&=&M_{i,j-1}\Theta(j-1) \longrightarrow \mathrm{shift\; right}
\label{zmmz}
\end{eqnarray}
where $\Theta(n)=1$ for $n>0$ and $\Theta(n)=0$ for $n\leq 0$.
When applied to a Toeplitz matrix $T_{jk}=t_{j-k}$ with $j,k=1,\ldots,n$, the operation $ZTZ^{\dag}$ shifts it along the main diagonal by one unit (hence the name displacement structure)
\begin{equation}
(Z\,T\, Z^{T})_{ij}=t_{i-j}\Theta(i-1)\Theta(j-1)\ .
\end{equation}
Therefore all terms in the displacement operator $\nabla_{Z, Z^T}(T)=T-Z\,T\,Z^{T}$ cancel except for $i,j=1$, and
\begin{equation}
 T-Z\,T\,Z^{T}=\left [ \begin{array}{ccccc}
t_0 & t_{-1} & t_{-2} & \ldots & t_{1-n}\\
t_1 & 0 & 0 & \ldots& 0\\
 \vdots&\ddots& \ddots& \ddots& \vdots\\
 t_{n-2} &0 & \ldots&0&0\\
 t_{n-1}& 0& \ldots& 0 &0 \end{array} \right ] \ . 
 \end{equation}
 Consequently, the displacement rank  for any Toeplitz matrix  is at most  $2$.
 
 For  Hankel matrices, $H_{jk}=h_{j+k}$ with $j,k=1,\ldots,n$ it is convenient to use the operation
\begin{equation}
(Z\,H-H\,Z^{T})_{ij}= h_{i-1+j}\Theta(i-1)-h_{i+j-1}\Theta(j-1).
\end{equation}
All terms in the displacement operator $\Delta_{Z,Z^{T}}(H)=Z\,H-H\,Z^{T}$ cancel except for $i,j=1$, and
\begin{equation}
Z\,H-H\,Z^{T} =\left [ \begin{array}{ccccc}
0 &- h_{2} & -h_{3} & \ldots & -h_{n}\\
h_2 & 0 & 0 & \ldots& 0\\
 \vdots&\ddots& \ddots& \ddots& \vdots\\
 h_{n-1} &0 & \ldots&0&0\\
 h_{n}& 0& \ldots& 0 &0 \end{array} \right ]\ .
 \label{displacement_Hankel}
 \end{equation}
Therefore the displacement rank of a Hankel matrix is also  at most $2$. 

For a Toeplitz-plus-Hankel matrix, $(T+H)_{ij}=t_{i-j}+h_{i+j}$ with $j,k=1,\ldots,n$, Eq.~\eqref{zmmz} yields the identities
 \begin{eqnarray}
(Z(T+H))_{ij}&=&(t_{i-j-1}+h_{i+j-1})\Theta(i-1) ,\\
(Z^{T}(T+H))_{ij}&=&(t_{i-j+1}+h_{i+j+1})\Theta(n-i)\\ 
 ((T+H)Z)_{ij}&=&(t_{i-j-1}+h_{i+j+1})\Theta(n-j) \\
 ((T+H)Z^{T})_{ij}&=&(t_{i-j+1}+h_{i+j-1})\Theta(j-1).
 \end{eqnarray}
Defining the displacement operator by \cite{heinig}
 \begin{equation}
 \Delta_{A,A}(T+H)=A(T+H)-(T+H)A,\qquad A=Z+Z^{T},
 \end{equation}
all terms in $\Delta_{A,A}(T+H)$ cancel except the boundary terms with $i=1,n$ and $j=1,n$, so that
it has the block structure
 \begin{equation}
\Delta_{A,A}(T+H)=
\left [ \begin{array}{ccc}
 t_1-t_{-1}                  &    -(t_{-j}+h_j)_{j=2}^{n-1}    & h_{2+n}-h_n   \\
 (t_i+h_i)_{i=2}^{n-1} &          (0)_{i,j=2}^{n-1}          & (t_{i-n+1}+h_{i+n-1} )_{i=2}^{n-1} \\
 h_{n}-h_{2+n} & -(t_{n-j-1}+h_{n+j+1})_{j=2}^{n-1}  & t_{-1}-t_1 \end{array} \right ].
\end{equation}
In general, the displacement rank of a Toeplitz-plus-Hankel matrix is at most $4$.

\section{Short-range plasma models} \label{plasma}

The usual semi-Poisson distribution corresponds to the case where only the nearest levels (from the ordered set)  interact by the factor $f(\lambda, \lambda^{\prime})=|\lambda^{\prime}-\lambda |^{\beta}$.
For this model all correlation functions are known \cite{gerland, atas}. $P_n(s)$ are gamma distributions with $\gamma_n=2n+1$ for $\beta=1$ and $\gamma_n=3n+2$ for $\beta=2$. The two-point correlation form factors  are given by the following formulas 
\begin{eqnarray}
K(\tau)&=&\frac{2+\pi^2 \tau^2}{4+\pi^2\tau^2},\qquad K(0)=\frac{1}{2}, \quad \beta=1, \\
K(\tau)&=&1-\frac{486}{729+108 \pi^2 \tau^2 +16 \pi^4 \tau^4}, \qquad   K(0)=\frac{1}{3},\quad \beta=2.
\end{eqnarray} 
 But  one can also consider the case when there exists also an interaction with next-to-nearest neighbors. Here we consider two such models. The first one has been discussed in detail in  \cite{gerland}. It corresponds to a model where the nearest and next-to-nearest levels interact by the same interaction: if $\lambda_1<\lambda_2<\lambda_3$ is any triple of nearest levels then they have the usual  interaction with $\beta=1$
\begin{equation}
|\lambda_3-\lambda_2| |\lambda_2-\lambda_1| |\lambda_3-\lambda_1|
\end{equation} 
It is easy to see that at small arguments the nearest-neighbor distributions have $s^{\gamma_n}$ behavior with  $\gamma_n=3n+1$.  The calculations in \cite{gerland} give
\begin{eqnarray}
P_0(s)&=&\frac{9}{2}(3-\sqrt{6}) s\exp(-3s)\Big (1+\frac{\sqrt{6}}{2}s\Big )^2 \nonumber\\
P_1(s)&=&\frac{3^4}{2^3}(5\sqrt{6}-12) s^4 \exp(-3s)\Big (1+\frac{\sqrt{6}}{2}s+\frac{3}{10}s^2\Big )
\label{k_2}\\
P_2(s)&=&\frac{3^8}{7\cdot 5 \cdot 2^5}(27-11\sqrt{6})s^7 \exp(-3s)\Big (1+\frac{11}{36}\sqrt{6}s+\frac{s^2}{8}\Big )\nonumber
\end{eqnarray}
The second model is chosen in such a way that the nearest levels interact with $\beta=2$ but the next-to-nearest levels interact with $\beta=1$
\begin{equation}
|\lambda_3-\lambda_2|^2 |\lambda_2-\lambda_1|^2 |\lambda_3-\lambda_1|
\end{equation} 
In this case 
\begin{equation}
P_n(s)\sim s^{\gamma_n},\qquad \gamma_n=4n+2.
\end{equation}
Such a case has not been considered in  \cite{gerland} but the  calculations can be done similarly as in the first model and the results are the following
\begin{eqnarray}
P_0(s)&=&2^5 (2-\sqrt{3})s^2\exp(-4s)\Big( 1+\frac{2\sqrt{3}}{3}s\Big)^2\nonumber\\
P_1(s)&=&\frac{2^{11}}{5\cdot 3^2}(7\sqrt{3}-12)s^6\exp(-4 s) \Big(1+\frac{2\sqrt{3}}{3}s+\frac{2}{7}s^2 \Big )\\
P_2(s)&=&\frac{2^{14} \cdot 13}{7\cdot 5^2\cdot 3^4}(26-15\sqrt{3})s^{10}\exp(-4s) 
\Big ( 1+\frac{60\sqrt{3}}{143}s+\frac{4}{33}s^2\Big )\,.
\end{eqnarray}
Similar formulas can be derived for the two-point correlation form factors. The direct calculations show that  the Laplace transform of the form factor has the following form
\begin{eqnarray}
g_2(t)&=&\frac{9}{4(3+t)}\left [ \frac{(6+t)^2}{(3+t)^3-3^3} +\frac{t^2(5-2\sqrt{6})}{(3+t)^3+3^3 (5-2\sqrt{6})}\right ] ,\qquad \gamma_n=3n+1,\label{g_first} \\ 
g_2(t)&=&\frac{16}{4+t}\left [ \frac{(8+t)^2}{(4+t)^4-4^4} +\frac{t^2(7-4\sqrt{3})}{(4+t)^4+4^4 (7-4\sqrt{3})}\right ], \qquad \gamma_n=4n+2.
\label{g_second}
\end{eqnarray}
The two-point form factor is related with the Laplace transform by \eqref{K_from_g}. The values of form factor at zero are  $1/3$ for the first model and $1/4$ for the second one, in accordance with results of Section~\ref{compressibility}. The corresponding curves are presented in Fig.~\ref{fig_formfactor}.


\end{document}